\documentclass[
prd
,showpacs ,amssymb,superscriptaddress,aps
]{revtex4}
\usepackage{graphicx}
\input{epsf}

\usepackage{amsmath,amssymb}
\usepackage{bm}
\usepackage{times}

\usepackage{aas_macros}

\newcommand{\dalm}{\kern1pt\vbox{\hrule height 0.9pt\hbox{\vrule width 0.9pt
\hskip 2.5pt\vbox{\vskip 5.5pt}\hskip 3pt\vrule width 0.3pt}\hrule height 0.3pt}
\kern1pt}

\newcommand{\gsim}{\, \raisebox{-0.8ex}{$\stackrel{\textstyle >}{\sim}$ }}

\begin{document}



\title{Gravitational wave asteroseismology with protoneutron stars}

\author{Hajime Sotani}
\email{hajime.sotani@nao.ac.jp}
\affiliation{Division of Theoretical Astronomy, National Astronomical Observatory of Japan, 2-21-1 Osawa, Mitaka, Tokyo 181-8588, Japan}

\author{Tomoya Takiwaki}
\affiliation{Division of Theoretical Astronomy, National Astronomical Observatory of Japan, 2-21-1 Osawa, Mitaka, Tokyo 181-8588, Japan}
\affiliation{Center for Computational Astrophysics, National Astronomical Observatory of Japan, 2-21-1 Osawa, Mitaka, Tokyo 181-8588, Japan}

\date{\today}

\begin{abstract}
We examine the time evolution of the frequencies of the gravitational wave after the bounce within the framework of relativistic linear perturbation theory using the results of one dimensional numerical simulations of core-collapse supernovae. Protoneutron star models are constructed in such a way that the mass and radius of protoneutron star become equivalent to the results obtained from the numerical simulations. Then, we find that the frequencies of gravitational waves radiating from protoneutron stars strongly depend on the mass and radius of protoneutron stars, but almost independently of the profiles of electron fraction and entropy per baryon inside the star. Additionally, we find that the frequencies of gravitational waves can be characterized by the square root of the average density of protoneutron star irrespectively the progenitor models, which are completely different from the empirical formula for cold neutron stars. The dependence  of the spectra on the mass and radius is different from that of the $g$-mode: the oscillations around the surface of protoneutron stars due to the convection and the standing accretion-shock instability.  Careful observations of the these modes of gravitational waves can determine the evolution of the mass and radius of protoneutron stars  after core-bounce. Furthermore, the expected frequencies of gravitational waves are around a few hundred hertz in the early stage after bounce, which must be a good candidate for the ground-based gravitational wave detectors. 
\end{abstract}

%
\maketitle
\section{Introduction}
\label{sec:I}

At the last moment of life of massive stars, neutron stars could be born in conjunction with supernova explosions. Neutron stars can realize extreme circumstances \cite{Haen2007}. The density inside a neutron star becomes easily over the nuclear saturation density. The magnetic and gravitational fields around/inside a neutron star are quite strong. Thus, neutron stars are good candidates to probe the physics under such extreme conditions. In practice, observations of neutron stars itself and/or of phenomena associated with compact objects would enable us to obtain an imprint of properties, which are difficult to provide on the Earth. For example, the discoveries of $2M_{\odot}$ neutron stars exclude some of soft equations of state (EOS) for neutron star matter \cite{D2010,A2013}. As another possibility, asteroseismology must be a powerful technique to extract the interior information of neutron stars via the observations of stellar oscillations, which are the same as seismology on the Earth and helioseismology on the Sun. Up to now, it has been suggested that one might get the information about the stellar mass, radius, EOS, and magnetic properties of compact objects through the oscillation spectra of gravitational waves (e.g., \cite{AK1996,AK1998,STM2002,SH2003,SYMT2011,PA2012,DGKK2013}). Furthermore, after the discoveries of the quasi-periodic oscillations in the afterglow of giant flares observed from soft-gamma repeaters (see Ref. \cite{WS2006} for a review), several attempts have been done especially to extract the EOS parameters by identifying the observed quasi-periodic oscillations with the crustal torsional oscillations \cite{SW2009,GNJL2011,SNIO2012,SNIO2013a,SNIO2013b,SIO2016}.

However, these examinations with the asteroseismological approach are almost restricted on the cases for cold neutron stars. That is, there are very few attempts for the protoneutron stars (PNSs) produced just after supernovae from the asteroseismological point of view. The EOS for supernova matter, where one has to consider the dependences of electron fraction and entropy, is quite different from that of neutron star matter, which is constructed under the beta equilibrium conditions. Therefore, one might obtain different aspects of EOS from the observations of oscillation spectra radiated from the PNSs. In fact, it has been pointed out that the oscillation properties of PNSs are quite different from those of cold neutron stars \cite{Ferrari:2002ut}, which is based on the models of isolated PNSs without mass accretion from the outer layer of the progenitors \cite{Pons:1998mm, Pons:2001ar}.

To hear the last voice of the massive stars, the next-generation gravitational wave detectors will be online in the coming years \cite{LIGO,VIRGO,KAGRA}.
The prediction of the gravitational wave from the next galactic supernova should be urgent task in order not to miss it \cite{OOGAGS2013,GSSZGO2016,NHTHTK2016,HKKT2015,TargetedSearch}. Recently numerical studies of core-collapse supernovae become highly sophisticated (see \cite{thierry15,tony15,JMS2016,burrows13,Kotake12_ptep} for reviews). With the multidimensional simulations, the imprint of the PNS oscillations have been found, which could be associated with the convection in the region around the surface of PNS \cite{MOB2009,MJM2013,Pablo2013,YMMYBHLBEVL2016}. Different from well-studied gravitational waves from rotational core-bounce that should be rare event \cite{DOJMM2007, HDKMRSY2008, SBFO2008, RBCDHM2009, LOHKS2012, AGDO2014, EMC2014, EFO2014, YAKSKKV2015} , this gravitational waves is emitted from ordinary non-rotating or slowly rotating PNS that should be majority of the remnants of core-collapse supernovae \cite{Langer2012}.

These oscillations correspond to the so-called $g$-modes, which is the specific oscillations with the restoring force due to the buoyancy. According to the literature \cite{MOB2009,MJM2013,Pablo2013,YMMYBHLBEVL2016}, the spectra of such $g$-modes oscillations from the PNSs are characterized with the properties of $\sim M_{\rm PNS}/R_{\rm PNS}^2$, where $M_{\rm PNS}$ and $R_{\rm PNS}$ denote the mass and radius of PNSs. Meanwhile, perturbative approaches might be better to extract the information of stellar oscillations rather than the direct numerical simulations. In the perturbative approaches, one can easily identify the physical background from the numerical calculations. Additionally, one can numerically calculate with high resolution owing to the low computational cost. So, we will consider to extract the oscillation information by using the perturbative approaches.

In this paper, we systematically examine the oscillation spectra of the PNSs with the relativistic Cowling approximation, where the metric is fixed during the stellar oscillations. For this purpose, the PNS models are constructed with the assumption that the PNSs are hydrostatic at each time step. The stellar mass and radius for each time step are determined from the fitting formula derived with the results of the stellar mass and radius by one dimensional (1D) numerical simulations of supernova, where the function forms of fitting is adopted the same as in Ref. \cite{SKJM2006}. The distributions of the entropy per baryon and electron fraction inside the PNSs are reproduced to mimic the results of 1D simulations of ours and those obtained in Ref. \cite{Roberts2012}. On such PNS models, we solve the eigenvalue problems to determine the oscillation spectra at each time step after the bounce in core-collapse supernova. We adopt geometric units, $c=G=1$, where $c$ and $G$ denote the speed of light and the gravitational constant, respectively, and the metric signature is $(-,+,+,+)$ in this paper.

\section{Protoneutron Star Models}
\label{sec:II}

We make models of PNS by two steps. First we have performed 1D time-dependent simulations of core-collapse supernovae and determine the mass and radius of PNS  as a function of the time. After that we solve general relativistic hydrostatic equation using the mass and radius obtained by the simulation at given post bounce time. The method for the time-dependent simulation is described in Section \ref{sec:IIa} and the method to solve the hydrostatic structure is provided in Section \ref{sec:IIb}. We remark that the numerical simulation for determining the radius and mass of PNS is done in Newtonian, while the PNS models to examine the frequencies of gravitational waves are constructed in general relativity as mentioned below. Thus, the radius of PNS might be overestimated, because the PNS model in general relativity could become more compact than that in Newtonian.

In this paper, we especially adopt the EOS proposed by Lattimer and Swesty \cite{LS} (the so-called LS220) and by Shen \cite{Shen}. LS220 is based on the compressible liquid drop model with the incompressibility $K_0=220$ MeV and the slope of the symmetry energy at the saturation density $L=73.82$ MeV. On the other hand, Shen EOS is based on the relativistic mean field theory with the TM1 nuclear interaction, where $K_0=281$ MeV and $L=114$ MeV. Using  a model of $15M_\odot $ progenitor \cite{WW1995}, we discuss how EOS affects the evolution of PNS.

Additionally, in order to examine the dependence of the progenitor mass, $M_{\rm pro}$, we also adopt  $M_{\rm pro}/M_\odot =11.2$, $27$, and $40$ \cite{WHW2002} with LS220. Hereafter, we consider the stellar model with $M_{\rm pro}/M_\odot=15$ constructed with LS220 as a typical model in this paper (we refer to this model as LS220M15.0), which is compared with the other models.

\subsection{1D Simulations of Core-Collapse Supernovae}
\label{sec:IIa}

First, we briefly show the setup and results of 1D simulations of core-collapse supernovae. The mass and radius of PNS calculated in the simulation are used to construct a model of PNS at given post bounce time. The simulations are performed with elaborate neutrino radiation hydrodynamics code based on our previous work \cite{TKS2016,NTKT2015}. The code combines hydrodynamics solver of high-resolution shock capturing scheme with approximate Riemann solver of HLLE \cite{HLLE} with neutrino transport solver that employ the isotropic diffusion source approximation (IDSA) \cite{idsa}. Our spatial grids has a finest mesh spacing, $\mathrm{d}r_{\rm min} = 0.25$km, at the center and covers up to 5000km with 512 non equidistant radial zones. The gravitational potential is calculated by Newtonian mono-pole approximation.

In the part of neutrino transport, we solve energy dependent moment equations of neutrinos for each neutrino energy, where 20 energy bins are employed to span 0 -- 300 MeV. In this study, we slightly developed our code to treat all flavors of neutrino in the framework of IDSA scheme.
The neutrino input physics are based on Bruenn (1985) \cite{Bruenn85}.
The reaction considered in this study is tabulated in Table \ref{tab:neutrino}.
Detail of the implementation is given at the forthcoming paper.

\begin{table}[htbp]
\begin{center}
\leavevmode
\caption{Summary of neutrino-matter interaction considered in the simulations}
\begin{tabular}{ccc}
\hline\hline
Reaction & Reference & Reference Number\\
\hline
$\nu_e+n \rightleftharpoons e^{-}+p$ & Bruenn (1985) & \cite{Bruenn85}\\
$\bar{\nu}_e+p \rightleftharpoons e^{+}+n$ & Bruenn (1985) & \cite{Bruenn85}\\
$\bar{\nu}_e+A^{\prime} \rightleftharpoons e^{-}+A$ & Bruenn (1985) &\cite{Bruenn85}\\
$\nu+e^{\pm}\rightleftharpoons \nu+e^{\pm}$ & Mezzacappa \& Bruenn (1993)  & \cite{MB1993,RJ2002}\\
$\nu+A\rightleftharpoons \nu+A$ & Bruenn (1985) with ion-ion correction & \cite{Bruenn85,Horowitz1997}\\
$\nu+\bar{\nu} \rightleftharpoons e^{-}+e^{+}$ & Bruenn (1985) & \cite{Bruenn85}\\
$\nu+\bar{\nu} + N + N \rightleftharpoons N+N$ & Hannestad and Raffelt (1998) & \cite{HR1998} \\
\hline\hline
\end{tabular}
\label{tab:neutrino}
\end{center}
\end{table}

\begin{figure*}
\begin{center}
\includegraphics[width=0.6\linewidth]{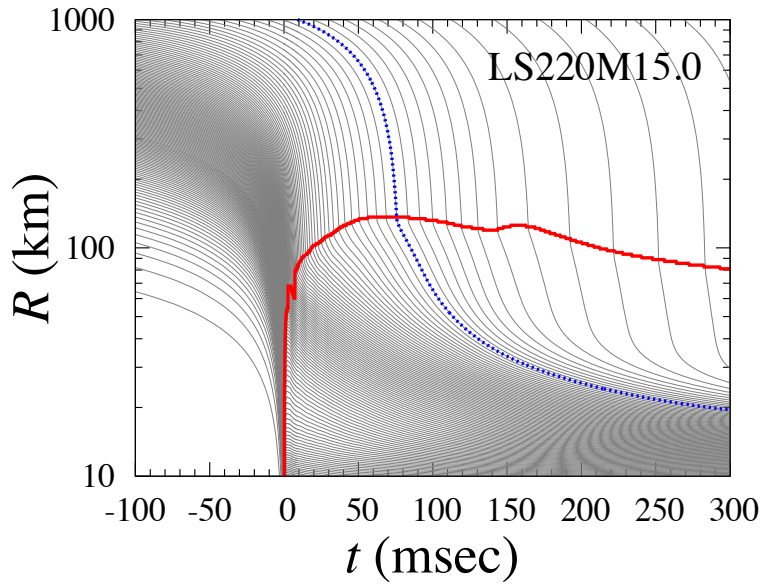} 
\end{center}
\caption{
Time evolutions of mass shell and shock radius of LS220M15.0. The position of the shock front is indicated by the thick-solid line. The evolution of mass shells are plotted by thin-solid lines ($\Delta M = 0.01 M_{\odot}$). In particular, the mass-shell whose enclosed mass is $1.4 M_\odot$ is plotted with the thick-dotted line.}
\label{fig:mass-shell}
\end{figure*}

Next we will show the overview of the results. 
The set of 15 $M_\odot$ progenitor of WW95 \cite{WW1995} and LS220 is widely used and well-tested in the context of supernova simulations. We select it to explain typical dynamics of the core-collapse and evolution of PNS.
The evolution of the shock and mass-shells are shown in Fig.\ref{fig:mass-shell}. 
The progenitor needs 175 ms to reach core-bounce.
At the core-bounce, the central value of $Y_e$ and $Y_l$ becomes $0.285$ and $0.34$, respectively,
where $Y_e$ and $Y_l$ are electron and lepton fractions.
The expansion of the shock front continues till 50 ms after bounce and the shock stalls at 140 km.
Our results shown above are roughly consistent with the feature of the previous work \cite{M2010}.

The shock does not revive in spite of the presence of  the neutrino heating,
since this is pure 1D simulation without employing the effect of the convection and other phenomenological enhancements of the neutrino heating.
It should be noted that our purpose of the simulation is to follow the evolution of PNS in the early stage, i.e., just 1 second after bounce.
Whether the shock revival happens or not is less important.

\begin{figure*}
\begin{center}
\begin{tabular}{cc}
\includegraphics[scale=0.5]{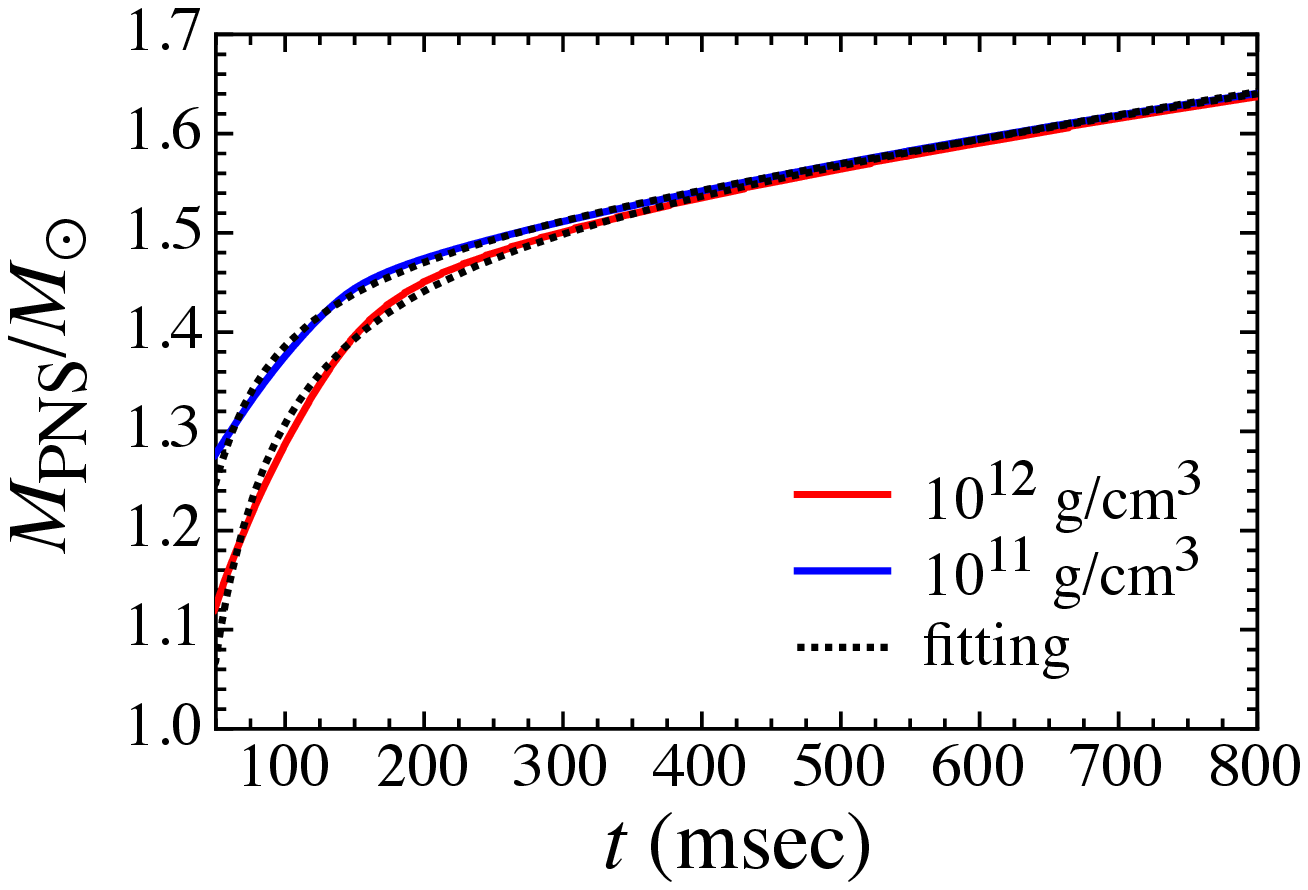} &
\includegraphics[scale=0.5]{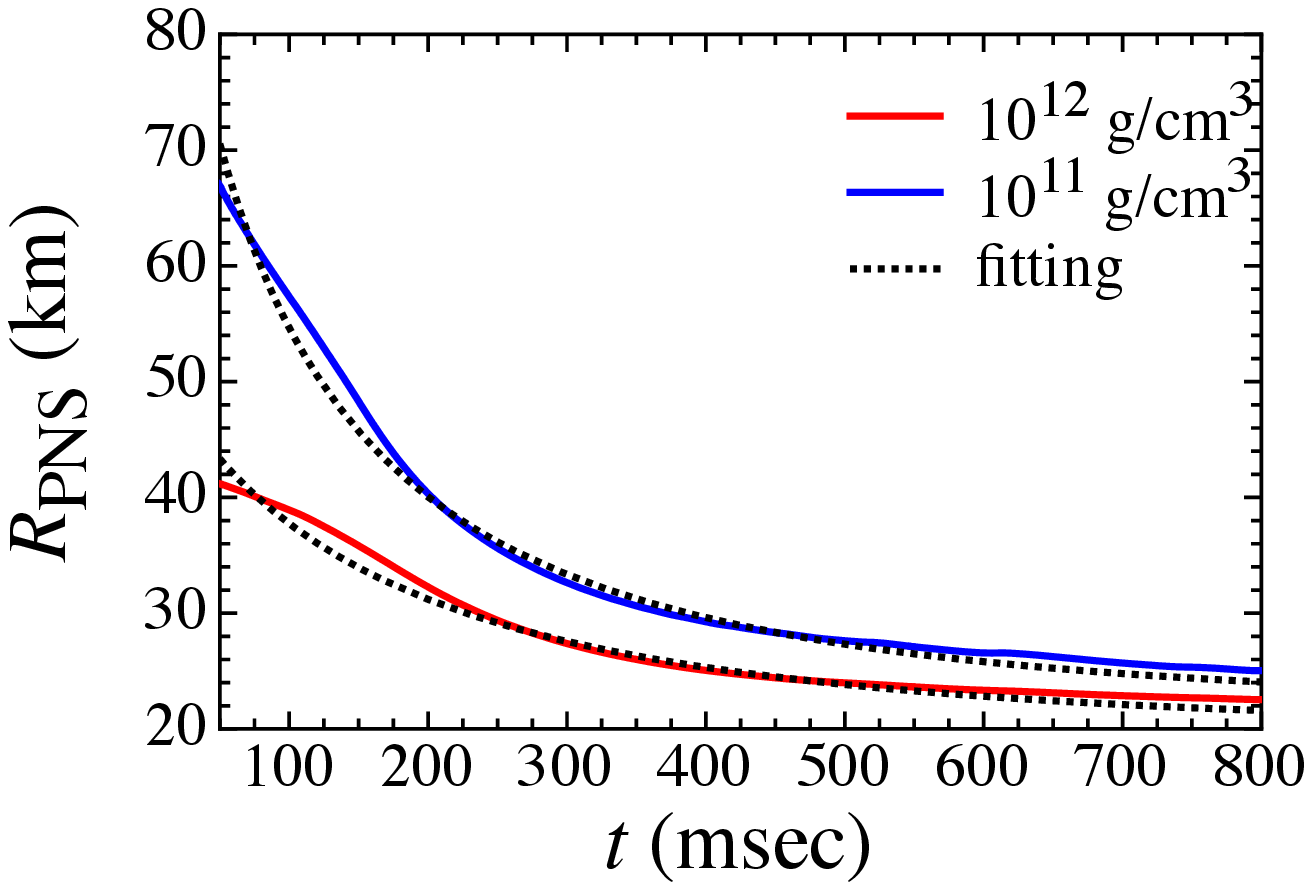}
\end{tabular}
\end{center}
\caption{
Time evolutions of $M_{\rm PNS}$ and $R_{\rm PNS}$ for the case of $M_{\rm pro}=15M_\odot$ and LS220, and its fitting with Eqs. (\ref{eq:Rt}) and (\ref{eq:Mt}). Two different lines correspond to the mass and radius of PNSs for the surface density with $10^{11}$ and $10^{12}$ g/cm$^3$, while the dotted lines denote the fitting to each case.  
}
\label{fig:MRLS220M15}
\end{figure*}

To describe PNS models, the information of the mass and radius of PNS is important. Fitting the numerical data with 1D simulation to the functional form proposed in Ref. \cite{SKJM2006}, we can get the fitting formula expressing the evolution of the PNS radius as
\begin{equation}
  R_{\rm PNS}(t) = \frac{R_{\rm i}}{1 + \left[1-\exp(-\frac{t}{\tau})\right]\left[\frac{R_{\rm i}}{R_{\rm f}} - 1\right]},   \label{eq:Rt}
\end{equation}
where $R_{\rm i}$ and $R_{\rm f}$ are the radii of the PNSs at $t=0$ and $\infty$, while $\tau$ denotes a typical time-scale for the evolution of the PNS in msec. We remark that $t=0$ refers to the time of bounce. In addition, we also obtain the fitting formula of the evolution of the PNS mass as
\begin{equation}
  \frac{M_{\rm PNS}(t)}{M_\odot} = \frac{c_0}{t} + c_1 t + c_2.   \label{eq:Mt}
\end{equation}
We note that $t$ in Eqs. (\ref{eq:Rt}) and (\ref{eq:Mt}) is in the unit of msec. The fitting parameters in Eqs. (\ref{eq:Rt}) and (\ref{eq:Mt}) for various stellar models are shown in Tables \ref{tab:Rt}, where the radius of PNS corresponds to the position that the density becomes $10^{12}$ g/cm$^3$, although the radius of PNS was sometimes considered to be the position that the density becomes $10^{11}$ g/cm$^3$. This point will be mentioned again where we show Fig. \ref{fig:Yes12}. For reference, we show the time evolution of mass and radius of PNSs for $M_{\rm pro}=15M_\odot$ and LS220, and its fitting in Fig. \ref{fig:MRLS220M15}, where two lines correspond to the PNSs for the surface density with $10^{11}$ and $10^{12}$ g/cm$^3$.

\begin{table*}[htbp]
\begin{center}
\leavevmode
\caption{Fitting parameters in the radius and mass evolutions of PNSs for various progenitor masses and EOSs. The stellar surface is determined where the density becomes $10^{12}$ g/cm$^3$.
}
\begin{tabular}{ccc|ccccccc  }
\hline\hline
  & EOS & $M_{\rm pro}/M_\odot$ & $R_{\rm i}$ (km) & $R_{\rm f}$ (km) & $\tau$ (msec) & $c_0$ (msec) & $c_1$ (msec$^{-1}$) & $c_2$ & \\
\hline
  & LS220 & 11.2 & 51.4 & 19.4 & 397 & $-6.629$ & $9.584\times 10^{-5}$ & $1.345$ & \\
  &            & 15.0 & 52.1 & 20.1 & 366 & $-22.86$  & $1.920\times 10^{-4}$ & $1.517$ & \\
  &            & 27.0 & 53.7 & 19.2 & 594 & $-34.86$ & $2.447\times 10^{-4}$ & $1.745$ & \\
  &            & 40.0 & 53.9 & 18.6 & 710 & $-39.97$ & $4.526\times 10^{-4}$ & $1.838$ & \\
\hline
  & Shen   & 15.0 & 48.4 & 21.1 & 559 & $-25.74$ & $2.463\times 10^{-4}$ & $1.498$ & \\
\hline\hline
\end{tabular}
\label{tab:Rt}
\end{center}
\end{table*}

\subsection{Modeling the Protoneutron Stars}
\label{sec:IIb}

\begin{figure}
\begin{center}
\includegraphics[scale=0.5]{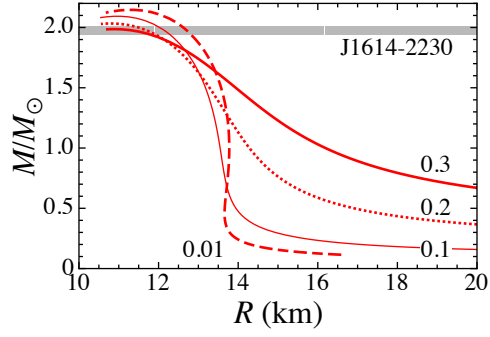}
\end{center}
\caption{
Mass-radius relation constructed with LS220 EOS for constant values of the entropy per baryon and $Y_e$ inside the star, where $s=1.5$ ($k_{\rm B}$/baryon) for various values of $Y_e=0.01$, $0.1$, $0.2$, and $0.3$. For reference, the mass of J1614-2230 \cite{D2010} is also shown in the figure. 
}
\label{fig:MRLS220}
\end{figure}

In order to mimic the structure of PNSs, we consider spherically symmetric objects, which are assumed to be hydrostatic in each moment of the evolution of PNSs. That is, PNS models in time are prepared in such a way that the masses and radii of a particular hydrostatic model are given by snapshots from the corresponding 1D hydrodynamic simulation. In fact, this might be a reasonable assumption as a first step, because the radial velocity of matter inside the PNS is sufficiently small. The metric describing the spherically symmetric objects is given by
\begin{equation}
  ds^2 =-e^{2\Phi} dt^2 + e^{2\Lambda} dr^2 + r^2\left(d\theta^2 + \sin^2\theta d\phi^2\right),
\end{equation}
where $\Phi$ and $\Lambda$ are the metric functions with respect to the radial coordinate $r$, and $\Lambda$ is associated with the mass function $m(r)$, such as $e^{-2\Lambda}=1-2m(r)/r$. The stellar models can be constructed by integrating the well-known Tolman-Oppenheimer-Volkoff equation together with the appropriate EOS. Unlike the case of cold neutron stars, where the stellar structure is determined under the beta equilibrium, to construct PNS models, one has to consider the distributions of entropy per baryon, $s$, and of $Y_e$ inside the star. For example, in Fig. \ref{fig:MRLS220}, we show the mass-radius relation constructed with LS220 EOS for constant values of $s$ and $Y_e$ inside the star, where $s$ is fixed to be $s=1.5$ ($k_{\rm B}$/baryon), while $Y_e$ is chosen to be $Y_e=0.01$, $0.1$, $0.2$, and $0.3$. From this figure, one can observe that the stellar radius for the fixed stellar mass increases with $Y_e$. In any case, one has to prepare the distributions of $s$ and $Y_e$ for constructing PNS models.

\begin{figure*}
\begin{center}
\begin{tabular}{cc}
\includegraphics[scale=0.5]{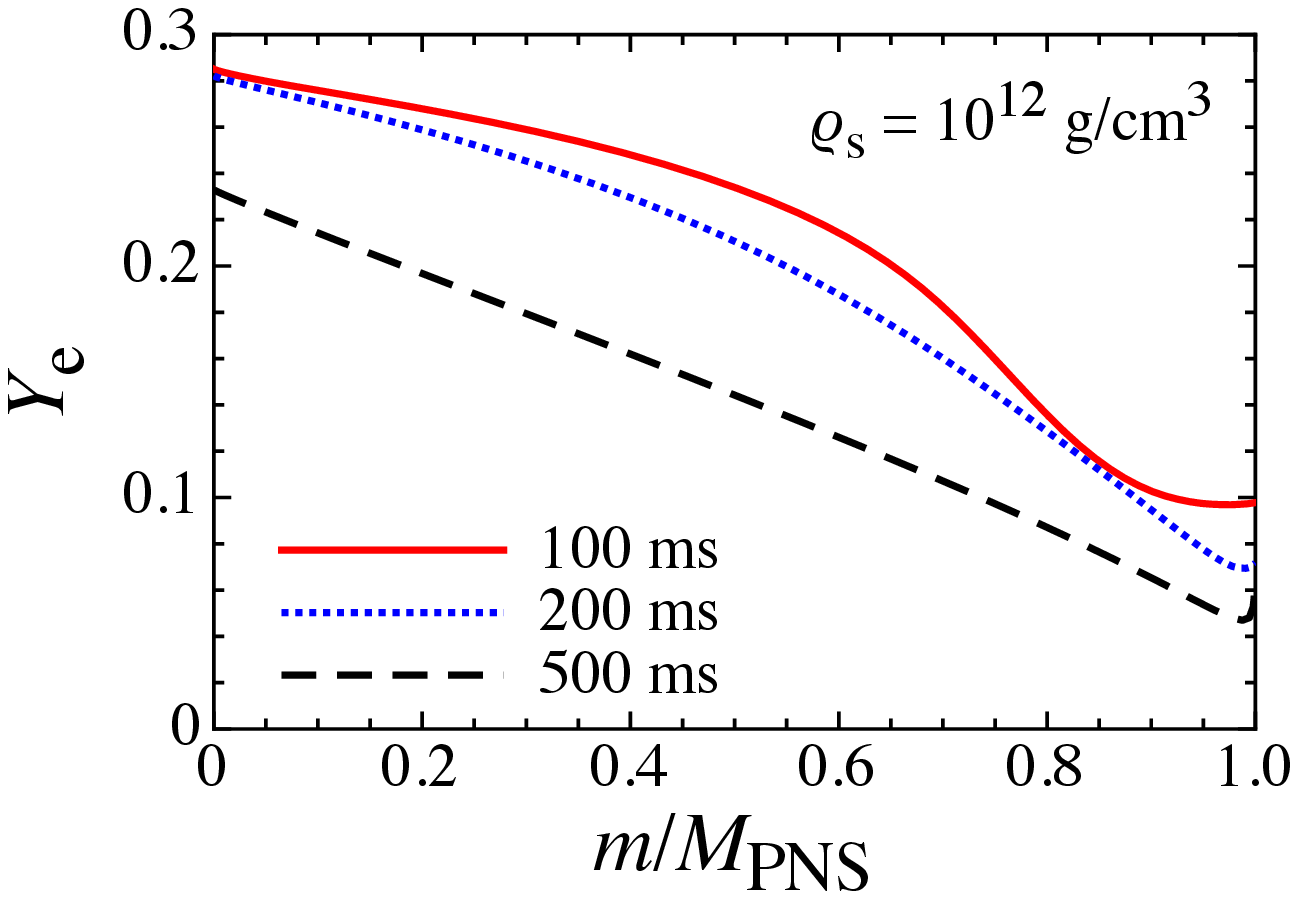} &
\includegraphics[scale=0.5]{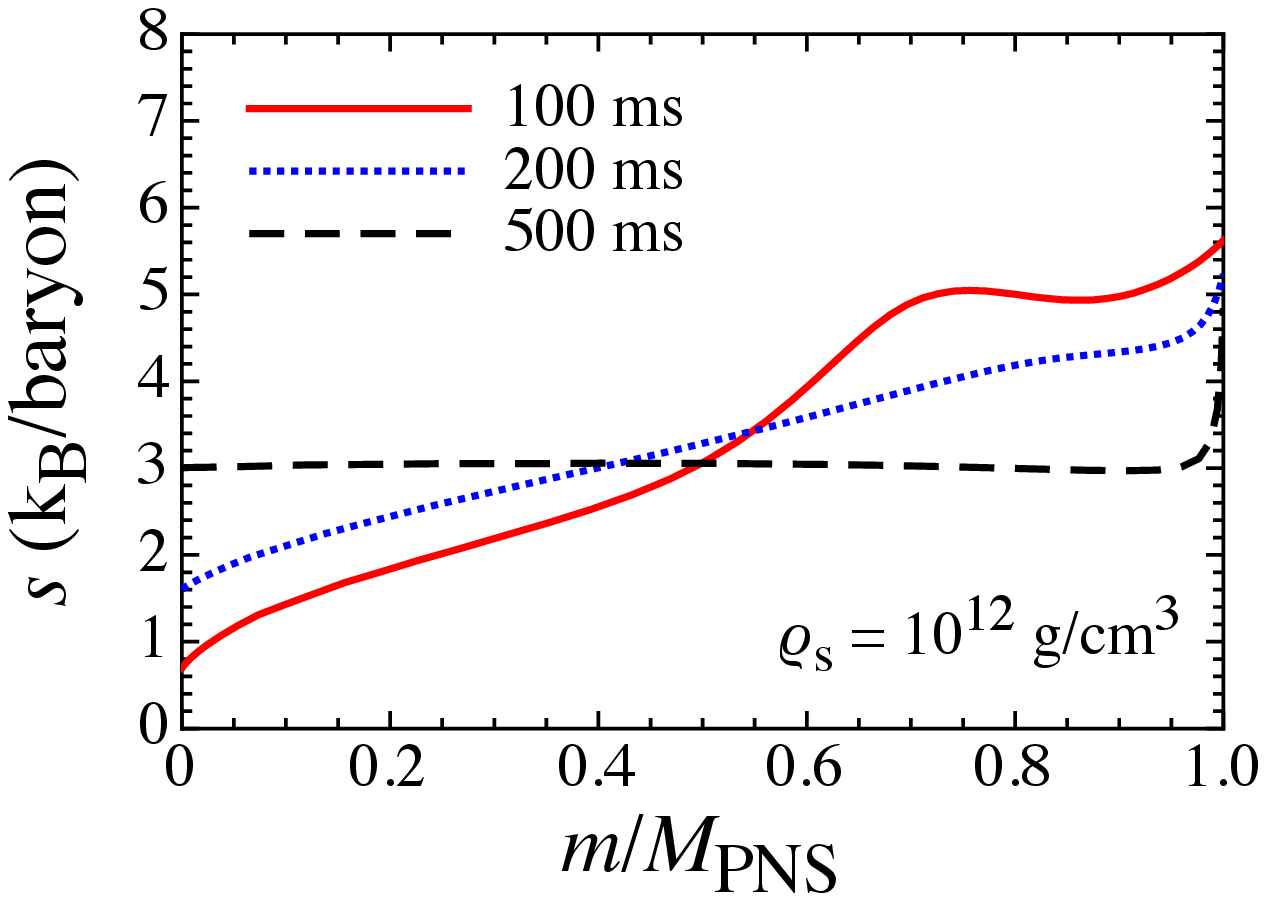}
\end{tabular}
\end{center}
\caption{
Distributions of electron fraction ($Y_e$) and entropy per baryon ($s$) as a function of mass coordinate normalized by the mass of PNS determined with $\rho_s=10^{12}$ g/cm$^3$. The different three lines correspond to the profiles of the different time after bounce, i.e., $t=100$, $200$, and $500$ ms.
}
\label{fig:Yes12}
\end{figure*}

\begin{figure*}
\begin{center}
\begin{tabular}{cc}
\includegraphics[scale=0.5]{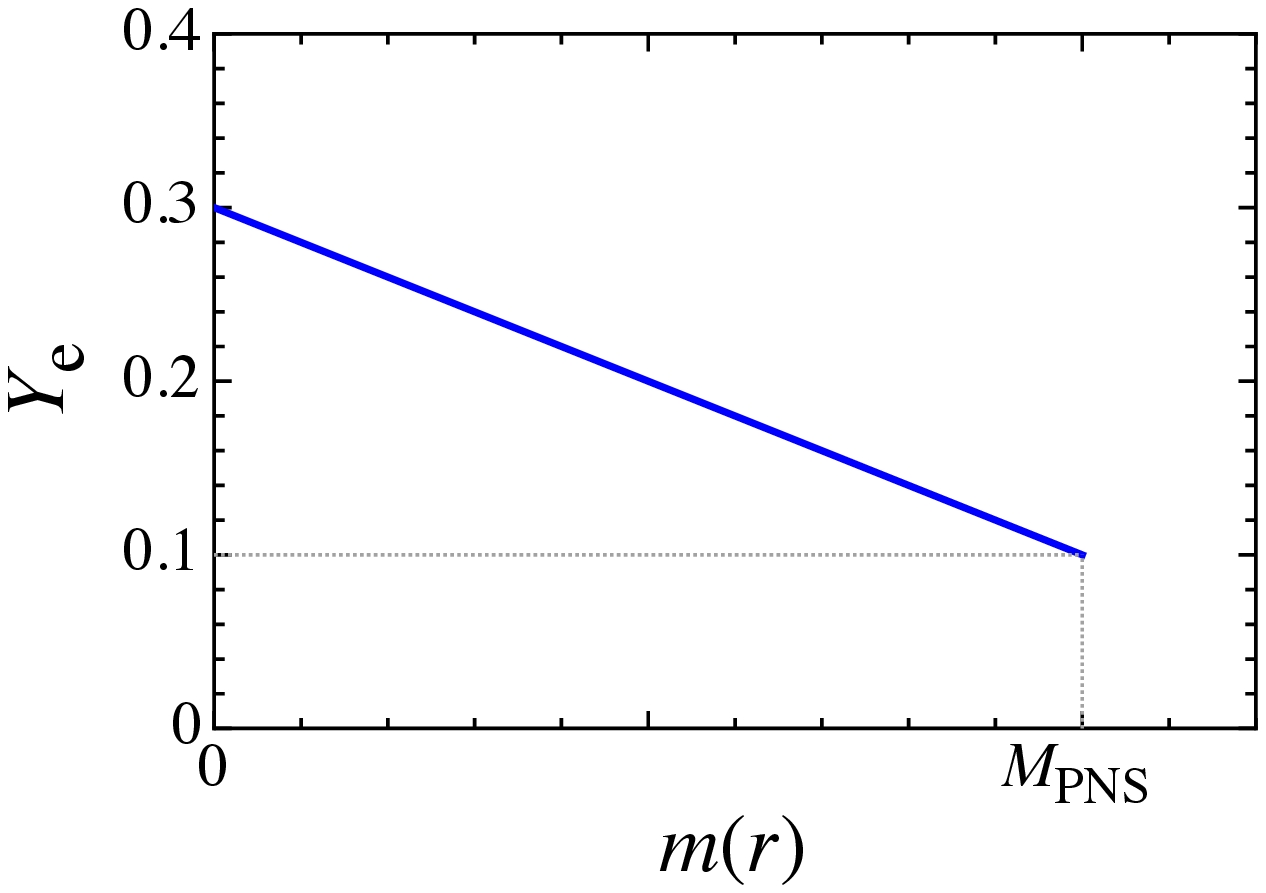} &
\includegraphics[scale=0.5]{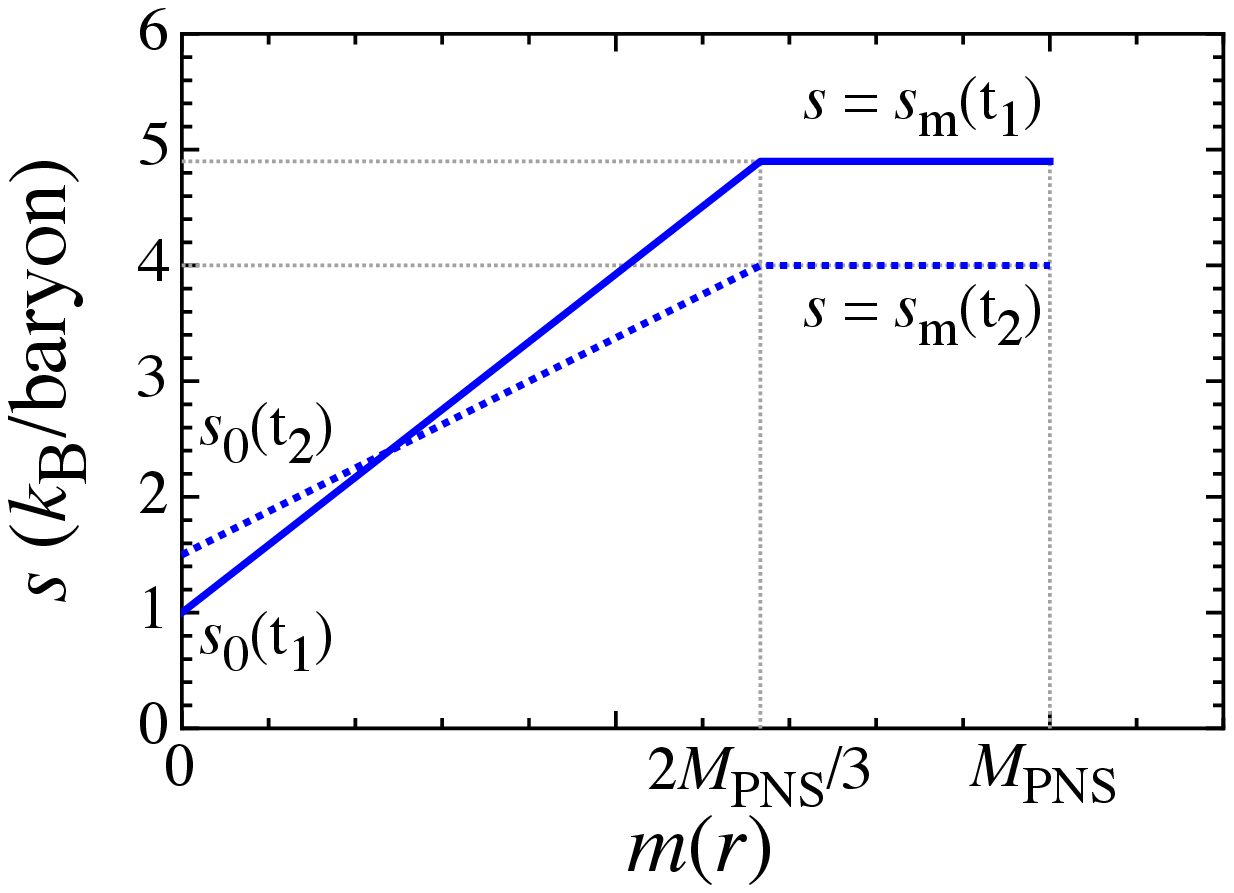}
\end{tabular}
\end{center}
\caption{
Distributions of the electron fraction $Y_e$ and entropy $s$ as a function of mass coordinate $m$, which simply imitates the results shown in Fig. \ref{fig:Yes12}. The distributions are expressed as Eqs. (\ref{eq:Ye2}) and (\ref{eq:s2}). In the right panel, $t_1$ and $t_2$ are evolution time, where $t_1<t_2$.
}
\label{fig:mYe}
\end{figure*}

For this purpose, we check the distributions of $s$ and $Y_e$ obtained from the 1D simulation mentioned in Sec. \ref{sec:IIa}. In Fig. \ref{fig:Yes12}, we show the $Y_e$ and $s$ distributions at $100$, $200$, and $500$ msec after bounce as a function of mass coordinate normalized by mass of PNSs \cite{entropy}, where the PNS surface is determined as the position that the density becomes $10^{12}$ g/cm$^3$. From this figure, one can see that $Y_e$ is roughly a linear function with negative slope, while $s$ is more complicated. To mimic these distributions, as shown in Fig. \ref{fig:mYe}, $Y_e$ and $s$ inside a PNS are simply assumed as
\begin{gather}
  Y_e = -0.2\bar{m} + 0.3,   \label{eq:Ye2} \\
  s =   \left\{
    \begin{array}{ll}
      3\bar{m}(s_m-s_0)/2 + s_0 &(0\le \bar{m}\le 2/3) \\
      s_m &(2/3\le \bar{m}\le 1) \ 
    \end{array}
  \right., \label{eq:s2}
\end{gather}
where $\bar{m} \equiv m/M_{\rm PNS}$, while $s_m$ is a constant for each time step. $s_0$ also depends on time, where we adopt the values of $s_0$ obtained from the 1D simulation at each time step. We remark that the profiles of $Y_e$ and $s$ dramatically change in the density region between $\sim 10^{11}$ and $10^{12}$ g/cm$^3$, where the values of $Y_e$ and $s$ abruptly increase. Then, the region for $\rho\simeq 10^{11} - 10^{12}$ g/cm$^3$ would  relatively be dilute, compared to the region for $\rho\gsim 10^{12}$ g/cm$^3$. So, to avoid such region, we choose the PNS surface is the position where the density becomes $10^{12}$ g/cm$^3$ in this paper.

\begin{figure*}
\begin{center}
\begin{tabular}{cc}
\includegraphics[scale=0.5]{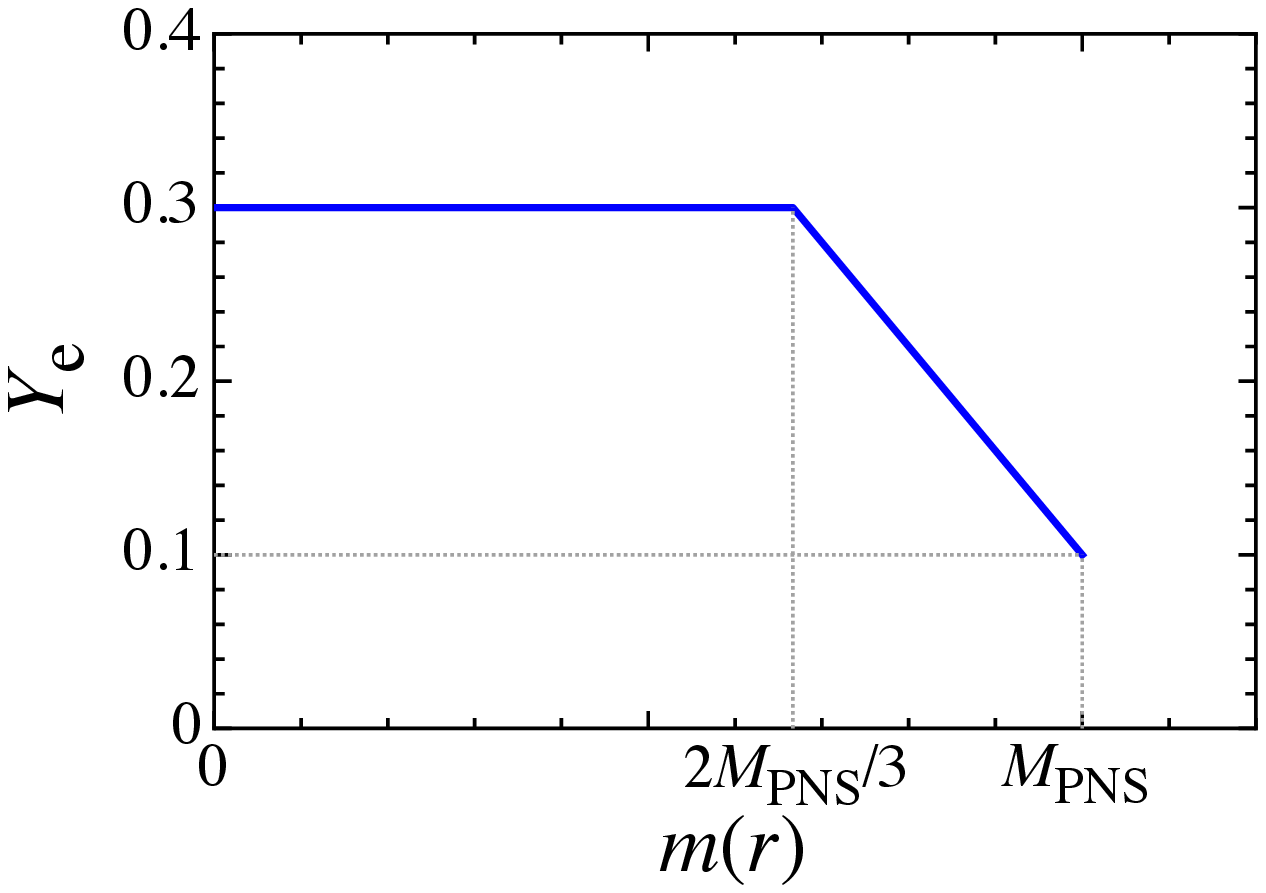} &
\includegraphics[scale=0.5]{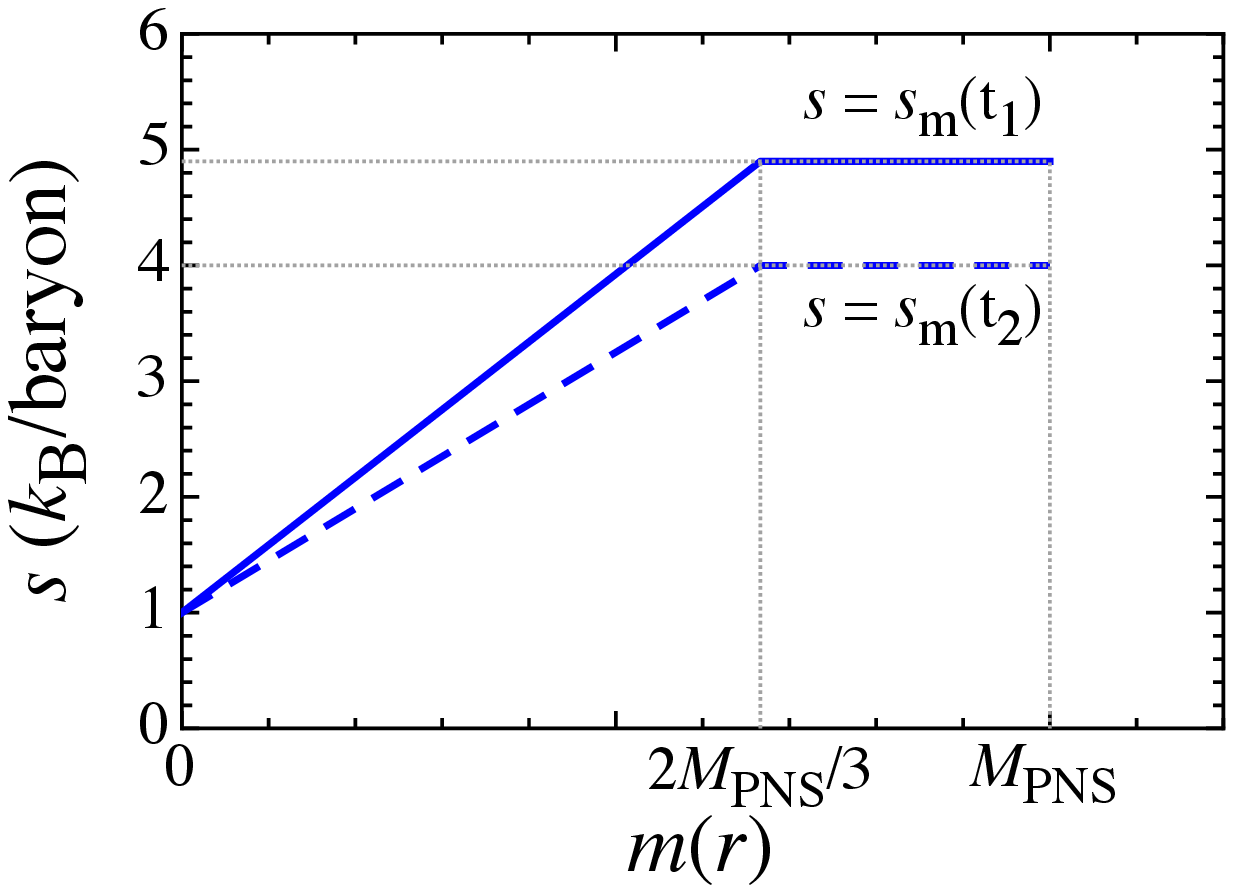}
\end{tabular}
\end{center}
\caption{
Distributions of the electron fraction $Y_e$ and entropy $s$ as a function of mass coordinate $m$, which simply imitates the results by Roberts \cite{Roberts2012}. The distributions are expressed as Eqs. (\ref{eq:Ye1}) and (\ref{eq:s1}). In the right panel, $t_1$ and $t_2$ are evolution time, where $t_1<t_2$.
}
\label{fig:mYe2}
\end{figure*}

On the other hand, the distributions of $Y_e$ and $s$ seem to be different from the results obtained by Roberts \cite{Roberts2012} (Fig. 1 in his paper), where he did 1D simulation of long-term evolution (for the first minute) of PNSs after bounce in general relativistic framework.  In order to examine how oscillation frequencies of PNSs depend on the distributions of $Y_e$ and $s$, we also consider an imitation inspired by the results of Roberts. That is, as shown in Fig. \ref{fig:mYe2}, $Y_e$ and $s$ distributions inside a PNS are expressed as
\begin{gather}
  Y_e =   \left\{
    \begin{array}{ll}
      0.3 &(0\le \bar{m}\le 2/3) \\
      -0.6\bar{m} + 0.7 &(2/3\le \bar{m}\le 1) \ 
    \end{array}
  \right., \label{eq:Ye1} \\
  s =   \left\{
    \begin{array}{ll}
      3\bar{m}(s_m-1)/2 + 1 &(0\le \bar{m}\le 2/3) \\
      s_m &(2/3\le \bar{m}\le 1) \ 
    \end{array}
  \right., \label{eq:s1}
\end{gather}
where $s_m$ is a constant for each time step, but could be different from the values determined in Eq. (\ref{eq:s2}).

\begin{figure*}
\begin{center}
\begin{tabular}{cc}
\includegraphics[scale=0.5]{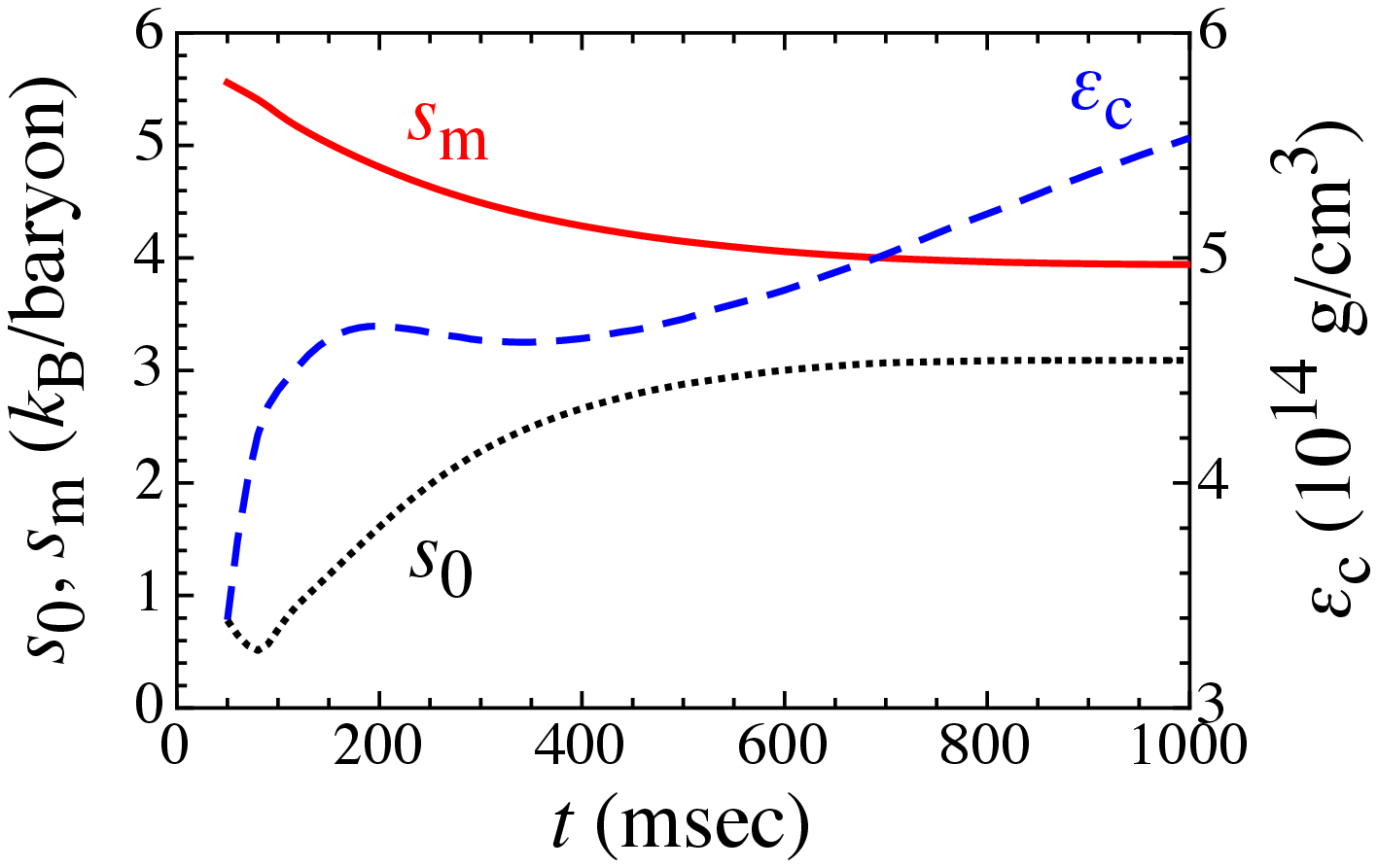} &
\includegraphics[scale=0.5]{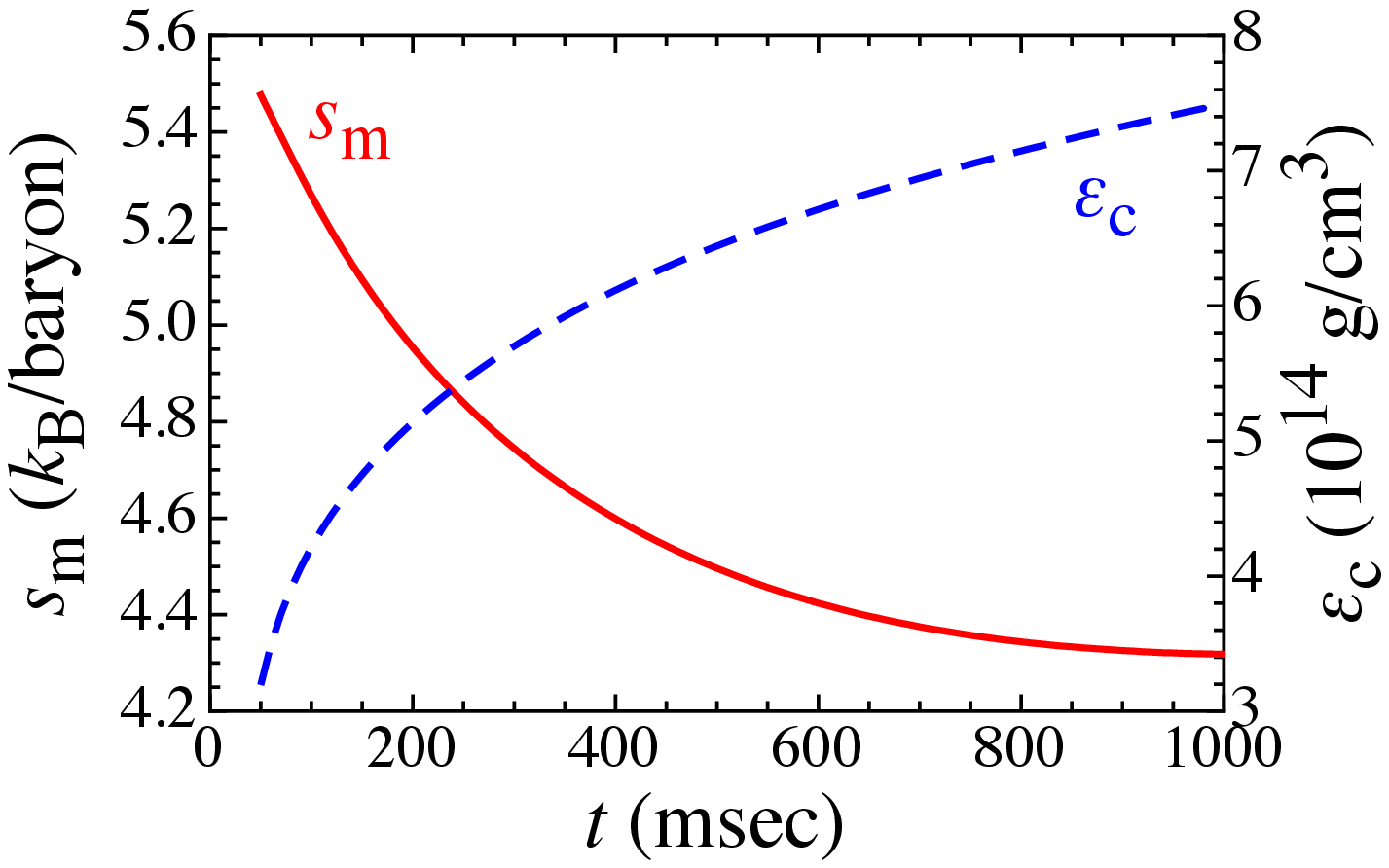}
\end{tabular}
\end{center}
\caption{
Values of $s_m$ and $\varepsilon_c$ for each time step to realize the mass and radius of PNSs for the case of $M_{\rm pro}=15M_\odot$ and LS220, where the left and right panels correspond to the combination of ($Y_e$, $s$) shown in Figs. \ref{fig:mYe} and \ref{fig:mYe2}, respectively. In left panel, the value of $s_0$ is also shown for reference.  
}
\label{fig:smt}
\end{figure*}

In both the combinations of ($Y_e$, $s$) mentioned above, there is one unknown parameter, $s_m$, while one has to choose a central energy-density, $\varepsilon_c$, to construct a PNS model. These two unknown values of $s_m$ and $\varepsilon_c$ are determined in such a way that the mass and radius of PNS constructed with $s_m$ and $\varepsilon_c$ should be equivalent to the given values of them for each time step, which are provided by Eqs. (\ref{eq:Rt}) and (\ref{eq:Mt}). In practice, for the case of $M_{\rm pro}=15M_\odot$ and LS220, the values of $s_m$ and $\varepsilon_c$ obtained with the combination of ($Y_e$, $s$) given as in Figs. \ref{fig:mYe} and \ref{fig:mYe2} are respectively shown as a function of time after bounce in left and right panels in Fig. \ref{fig:smt}. In left panel, we also show the the value of $s_0$, which is numerically obtained from the1D simulation. From this figure, one observes that the evolution of $s_m$ and $\varepsilon_c$ can strongly depend on the distributions of $Y_e$ and $s$ inside PNSs even for the same mass and radius of the PNS.

\section{Evolution of oscillation spectra}
\label{sec:III}

In order to examine oscillation spectra of PNSs, as a fist step, we adopt the relativistic Cowling approximation, where metric perturbations are neglected during stellar oscillations, i.e., $\delta g_{\mu\nu}=0$. In addition, for simplicity, we neglect an entropy variation in this paper, although one should generally consider it for the oscillations of PNSs. This is because the thermal energy (not entire internal energy) is still much smaller than the gravitational energy \cite{thermal}, even though the temperature of PNS is around a few tens of MeV \cite{Roberts2012}. On these assumptions, we consider polar type oscillations, which are strongly associated with the gravitational radiations, because the polar type oscillations involve density variations.

Considering polar parity oscillations of PNSs, the Lagrangian displacement vector of the fluid element is expressed as
\begin{equation}
  \xi^i = (e^{-\Lambda}W,-V\partial_\theta,-V\sin^{-2}\theta\partial_\phi)r^{-2}Y_{\ell m},
\end{equation}
where $W$ and $V$ are functions with respect to $t$ and $r$, while $Y_{\ell m}$ denotes the spherical harmonics function with the azimuthal quantum number $\ell$ and the magnetic quantum number $m$. So, the perturbed four velocity becomes
\begin{equation}
  \delta u^\mu = (0,e^{-\Lambda}\partial_t W, -\partial_t V\partial_\theta, -\partial_t V\sin^{-2} \theta \partial_\phi) r^{-2}e^{-\Phi}Y_{\ell m}.
\end{equation}
Then, one can derive the perturbation equations expressing the fluid oscillations by linearizing the energy-momentum conservation law, i.e., $\delta (\nabla_\nu T^{\mu\nu})=0$, which reduces to $\nabla_\nu \delta T^{\mu\nu}=0$ with the Cowling approximation. The explicit form of perturbation equations is written as 
\begin{align}
  &\frac{\varepsilon+p}{r^2}e^{\Lambda-2\Phi}\ddot{W} - \partial_r\left[\frac{\gamma p}{r^2}
     \left\{e^{-\Lambda}W' + \ell(\ell+1)V\right\}+e^{-\Lambda}p'\frac{W}{r^2}\right] \nonumber \\
  &\hspace{2cm} + \frac{p'}{r^2}\left(1+\frac{dp}{d\varepsilon}\right)\left[e^{-\Lambda}W'+\ell(\ell+1)V\right] 
     - \frac{\varepsilon' + p'}{r^2}\Phi'e^{-\Lambda}W = 0, \label{Tr} \\
  &(\varepsilon + p)e^{-2\Phi}\ddot{V} + \frac{\gamma p}{r^2}\left[e^{-\Lambda}W' + \ell(\ell+1)V\right] + \frac{p'}{r^2}e^{-\Lambda}W = 0, \label{Ttheta}
\end{align}
which correspond to $\nabla_\nu \delta T^{\mu\nu}=0$ for $\mu=r$ and $\theta$, respectively. In these equations, the dot and prime on the variables denote the partial derivatives with respect to $t$ and $r$, respectively, while $\gamma$ is the adiabatic index defined as
\begin{equation}
  \gamma \equiv \frac{\varepsilon + p}{p}\left(\frac{\partial p}{\partial \varepsilon}\right)_s. \label{gamma}
\end{equation}

The above perturbation equations can be simplified more by calculating the form [Eq. (\ref{Tr})]$+\partial_r$[Eq. (\ref{Ttheta})], which leads to
\begin{align}
  (\varepsilon' + p') e^{-2\Phi}\ddot{V} -& 2\Phi'(\varepsilon + p) e^{-2\Phi}\ddot{V} + (\varepsilon + p) e^{-2\Phi}\ddot{V}' \nonumber \\
     +& \frac{\varepsilon + p}{r^2} e^{\Lambda - 2\Phi} \ddot{W}  - \frac{\varepsilon' + p'}{r^2}\Phi' e^{-\Lambda}W  \nonumber \\
     +& \frac{p'}{r^2}\left(1 + \frac{dp}{d\varepsilon}\right) \left[e^{-\Lambda} W' + \ell(\ell+1)V\right]= 0.
\end{align}
From this equation, by removing the term of $W'$ with Eq. (\ref{Ttheta}), one can obtain the simple equation, such as
\begin{equation}
  \ddot{V}' = 2\Phi' \ddot{V} - \frac{1}{r^2}e^{\Lambda}\ddot{W}.  \label{eq:ddotV}
\end{equation}
Finally, assuming a harmonic dependence on time, such as $W(t,r)=W(r)e^{i\omega t}$ and $V(t,r)=V(r)e^{i\omega t}$, from Eqs. (\ref{Ttheta}) and (\ref{eq:ddotV}), the perturbation equations can be written as
\begin{gather}
  W' = \frac{\partial \varepsilon}{\partial p} \left[\Phi' W + \omega^2 r^2 e^{\Lambda - 2\Phi} V\right] - \ell(\ell+1)e^{\Lambda}V,    \label{eq:dW}  \\ 
  V' =  2\Phi' V - \frac{1}{r^2}e^{\Lambda}W. \label{eq:dV}
\end{gather}

Imposing appropriate boundary conditions, the problem to solve reduces to the eigenvalue problem with respect to the eigenvalue $\omega$. In this paper, we adopt the same boundary conditions as for determining the oscillation spectra of cold neutron stars \cite{SYMT2011}. That is, we impose the regularity condition at the stellar center and the condition that the Lagrangian perturbation of pressure should be zero at the surface of PNSs. From the regularity condition at $r=0$, the variables can be expressed as $W(r)=\alpha r^{\ell+1} + {\cal O}(r^{\ell + 3})$ and $V(r) = -\alpha r^\ell/\ell + {\cal O}(r^{\ell+2})$ in the vicinity of the stellar center, where $\alpha$ is an arbitrary constant. On the other hand, the boundary condition at the surface of PNS is given by \cite{SYMT2011}
\begin{equation}
 \omega^2r^2e^{\Lambda-2\Phi}V + \Phi' W = 0.
\end{equation}
We remark that the boundary condition adopted at the surface of PNSs might be inappropriate, since the matter still exists outside the PNSs. However, the density at the surface of PNSs abruptly drops down and the density outside the PNSs must be much smaller than that inside the PNSs. Thus, we believe that frequencies of PNSs could be qualitatively examined well even with such a boundary condition at the surface.

\begin{figure*}
\begin{center}
\begin{tabular}{ccc}
\includegraphics[scale=0.42]{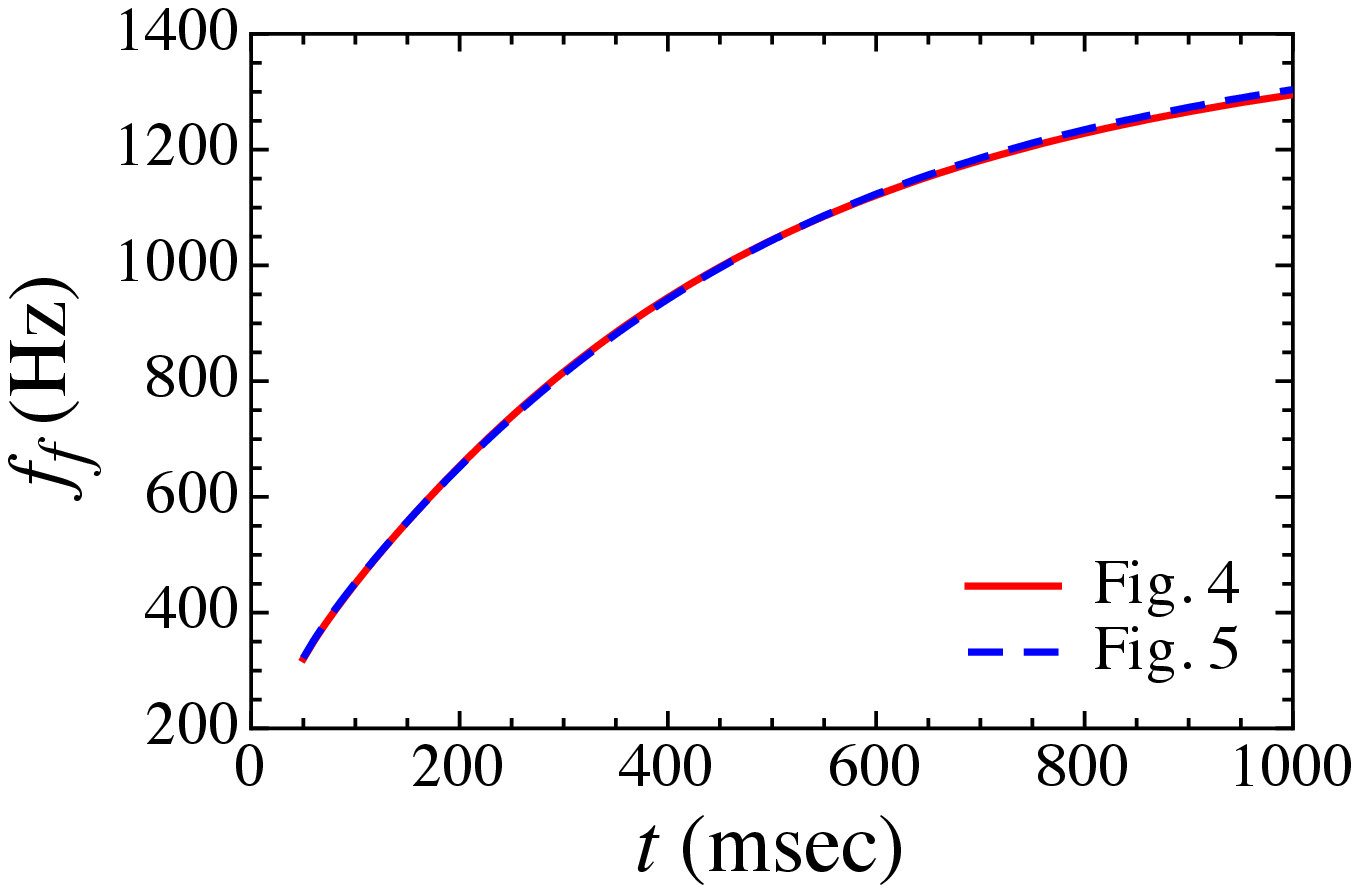} &
\includegraphics[scale=0.42]{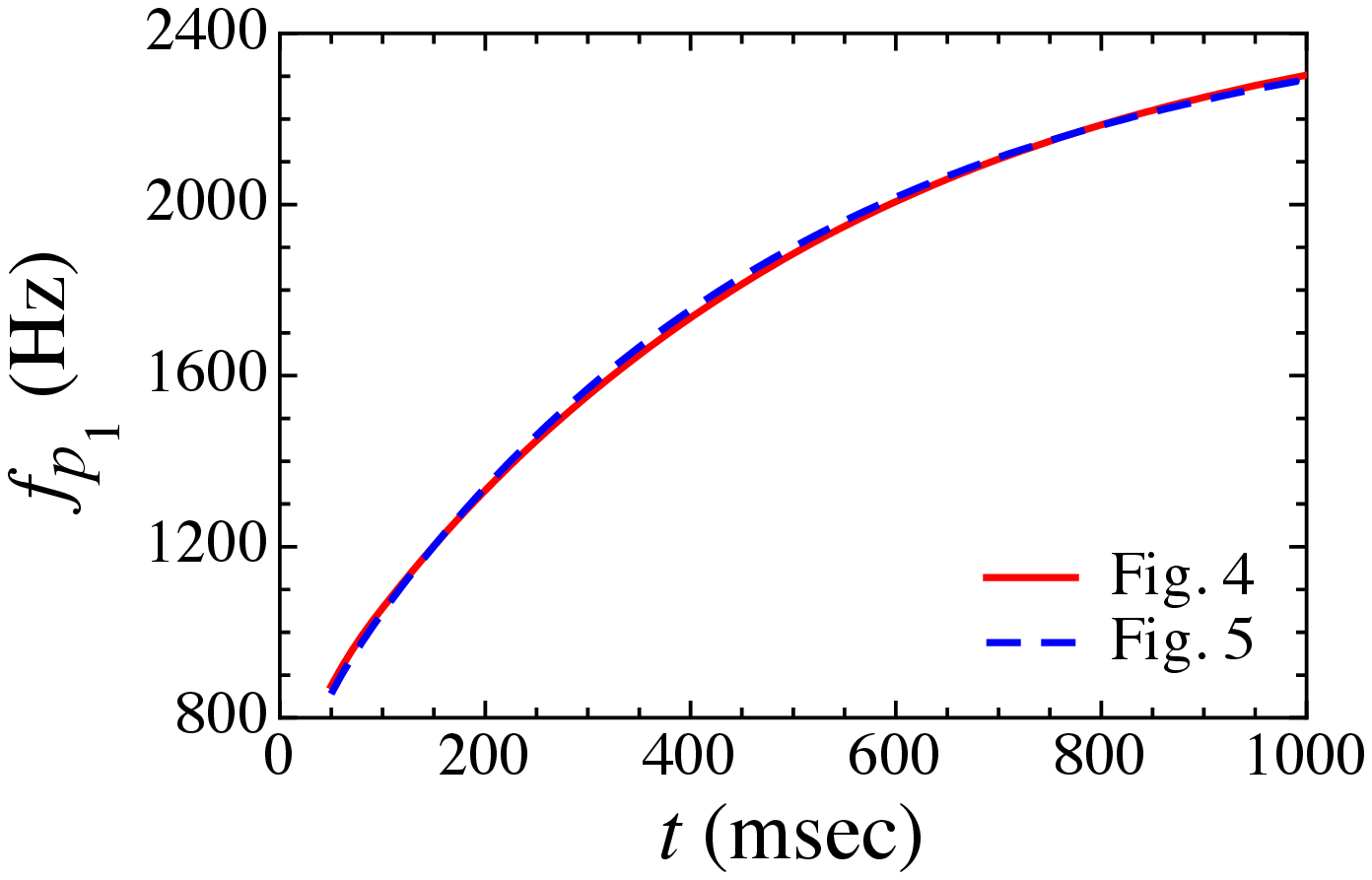} &
\includegraphics[scale=0.42]{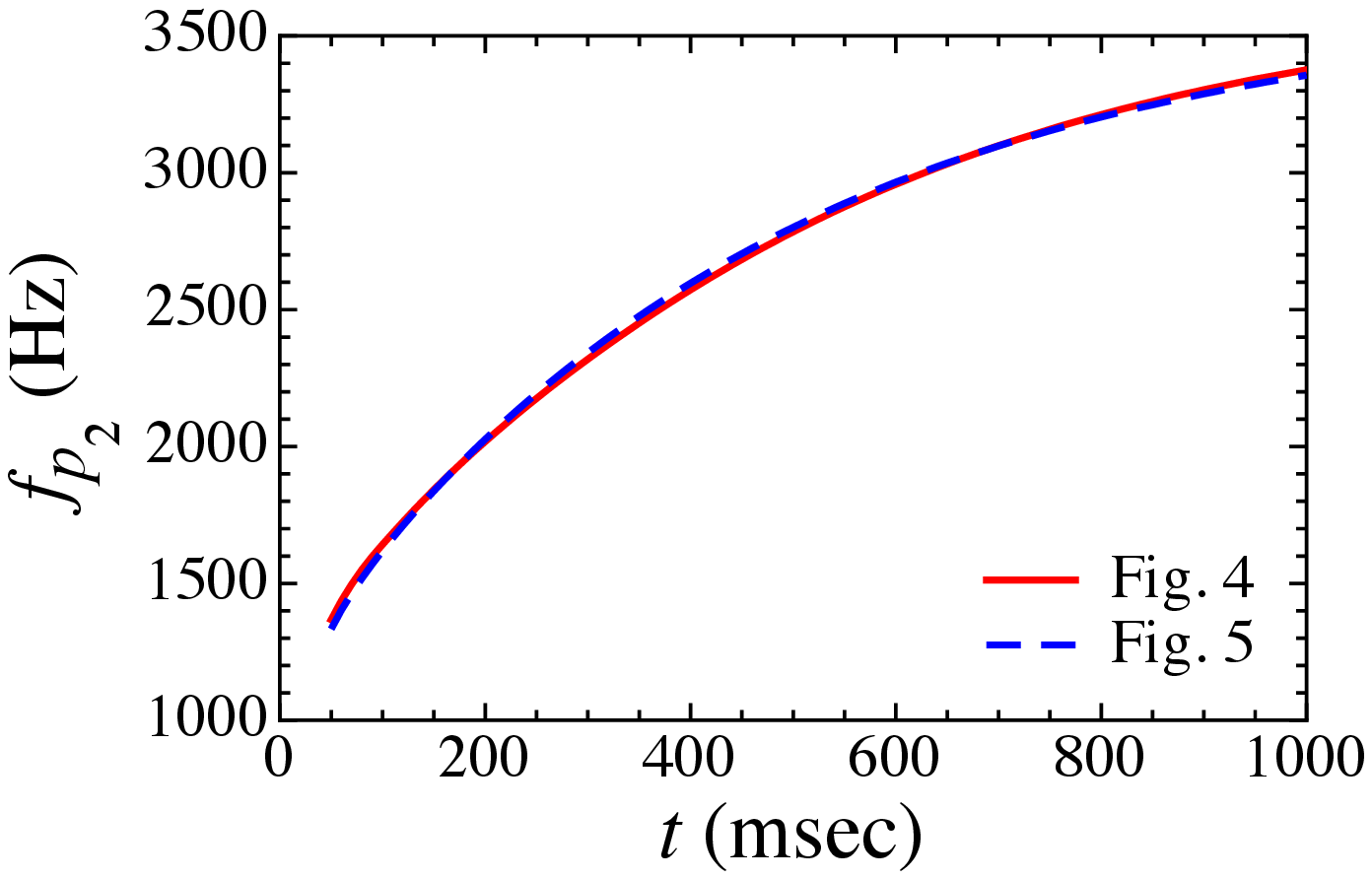}
\end{tabular}
\end{center}
\caption{
For the PNS model with $M_{\rm pro}=15M_\odot$ and LS220, time evolutions of eigenfrequencies are plotted, where the left, middle, and right panels correspond to $f$-, $p_1$-, and $p_2$-modes. In the figure, the different lines denote the results with different ($Y_e$, $s$) distributions inside the star, i.e., the solid and broken lines correspond to the distributions as shown in Figs. \ref{fig:mYe} and \ref{fig:mYe2}, respectively.
}
\label{fig:LS220M15}
\end{figure*}

\begin{figure*}
\begin{center}
\begin{tabular}{ccc}
\includegraphics[scale=0.42]{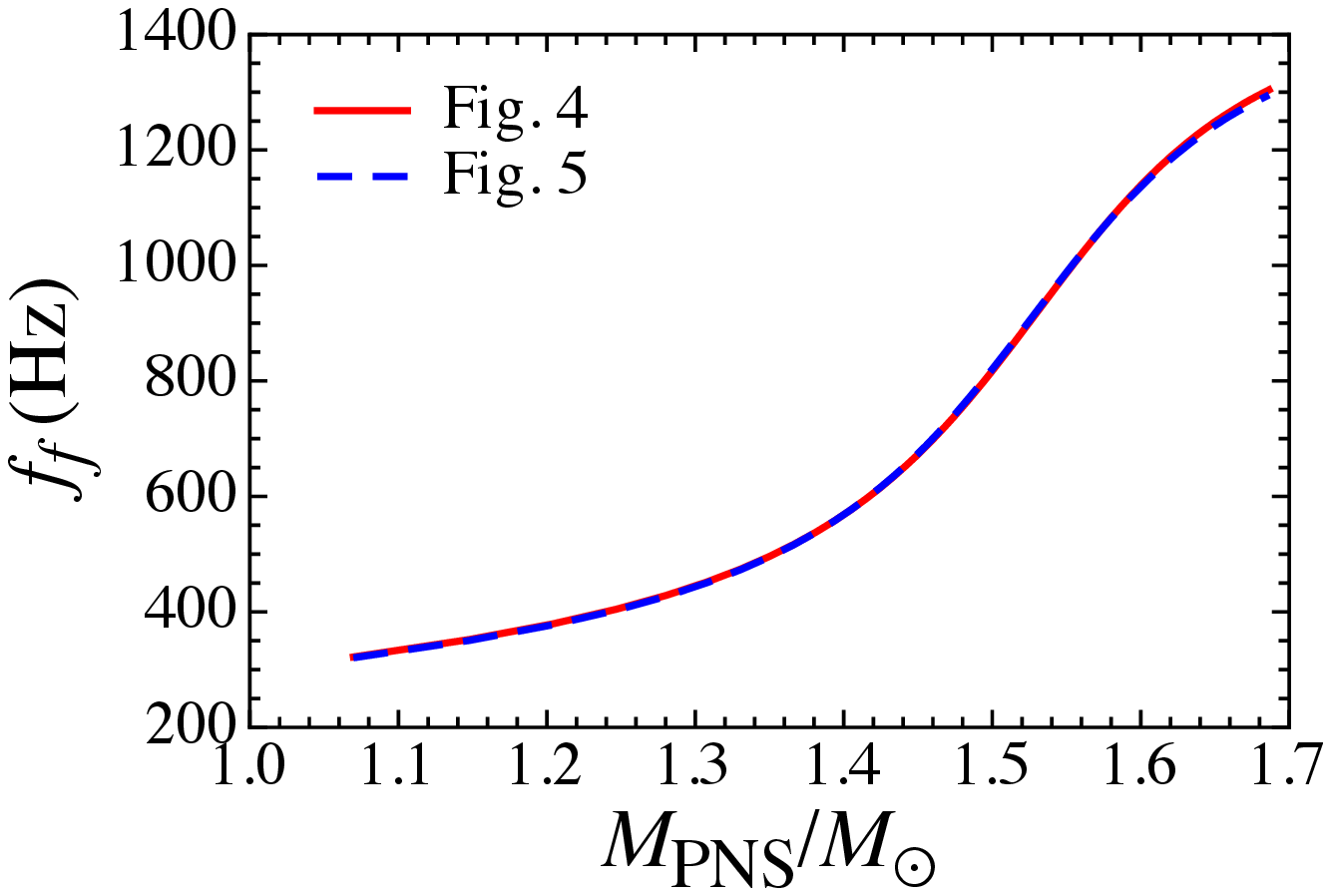} &
\includegraphics[scale=0.42]{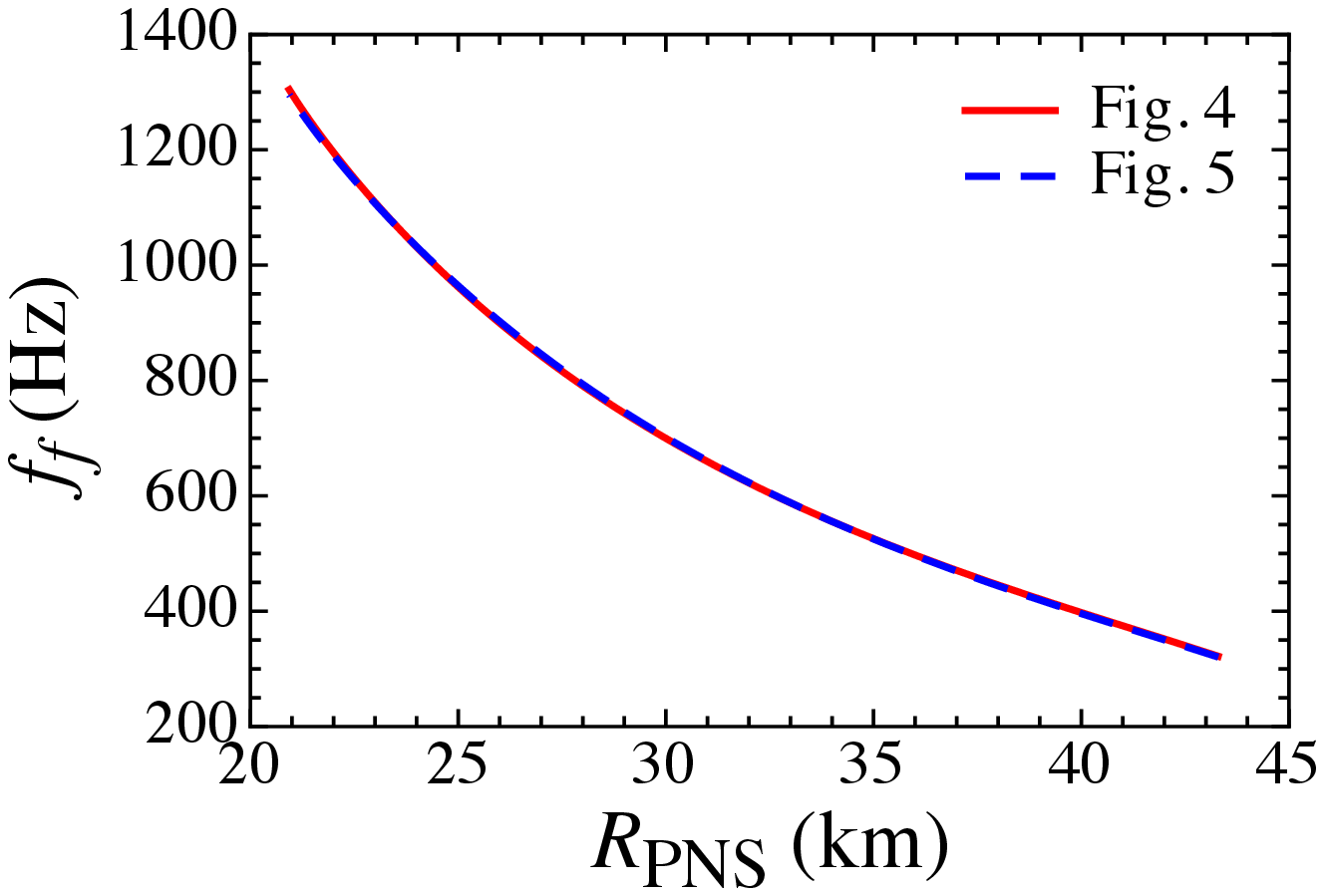}
\end{tabular}
\end{center}
\caption{
For the PNS model with $M_{\rm pro}=15M_\odot$ and LS220, the $f$-mode eigenfrequencies are plotted as a function of PNS mass (left panel) and radius (right panel). We remark that the PNS mass and radius are time-dependent as shown in Fig. \ref{fig:MRLS220M15}. As in Fig. \ref{fig:LS220M15}, the different lines denote the results with different ($Y_e$, $s$) distributions inside the star.
}
\label{fig:ff-LS220M15}
\end{figure*}

First, in Fig. \ref{fig:LS220M15}, we show the time evolutions of frequencies of $f$-, $p_1$-, and $p_2$-modes for the PNS model with $M_{\rm pro}=15M_\odot$ and LS220, where the solid and broken lines correspond to the frequencies obtained with the ($Y_e$, $s$) distributions inside the star shown in Figs. \ref{fig:mYe} and \ref{fig:mYe2}, respectively. From this figure, we can find that the dependence of the frequencies on the ($Y_e$, $s$) distributions seems to be weak. On the other hand, since the dependence of the frequencies on time comes from the different stellar mass and radius of PNS model for each time step as shown in Fig. \ref{fig:MRLS220M15}, the frequencies depend strongly on the PNS mass and radius. In order to clarify such a dependence, the $f$-mode frequency is shown as a function of the PNS mass and radius in Fig. \ref{fig:ff-LS220M15} with the different ($Y_e$, $s$) distributions inside the star. We remark that the frequencies of $f$-mode oscillations are predicted around a few hundred of hertz at the early stage after bounce, which could be very good signal for ground-based gravitational wave detectors.

\begin{figure*}
\begin{center}
\begin{tabular}{ccc}
\includegraphics[scale=0.42]{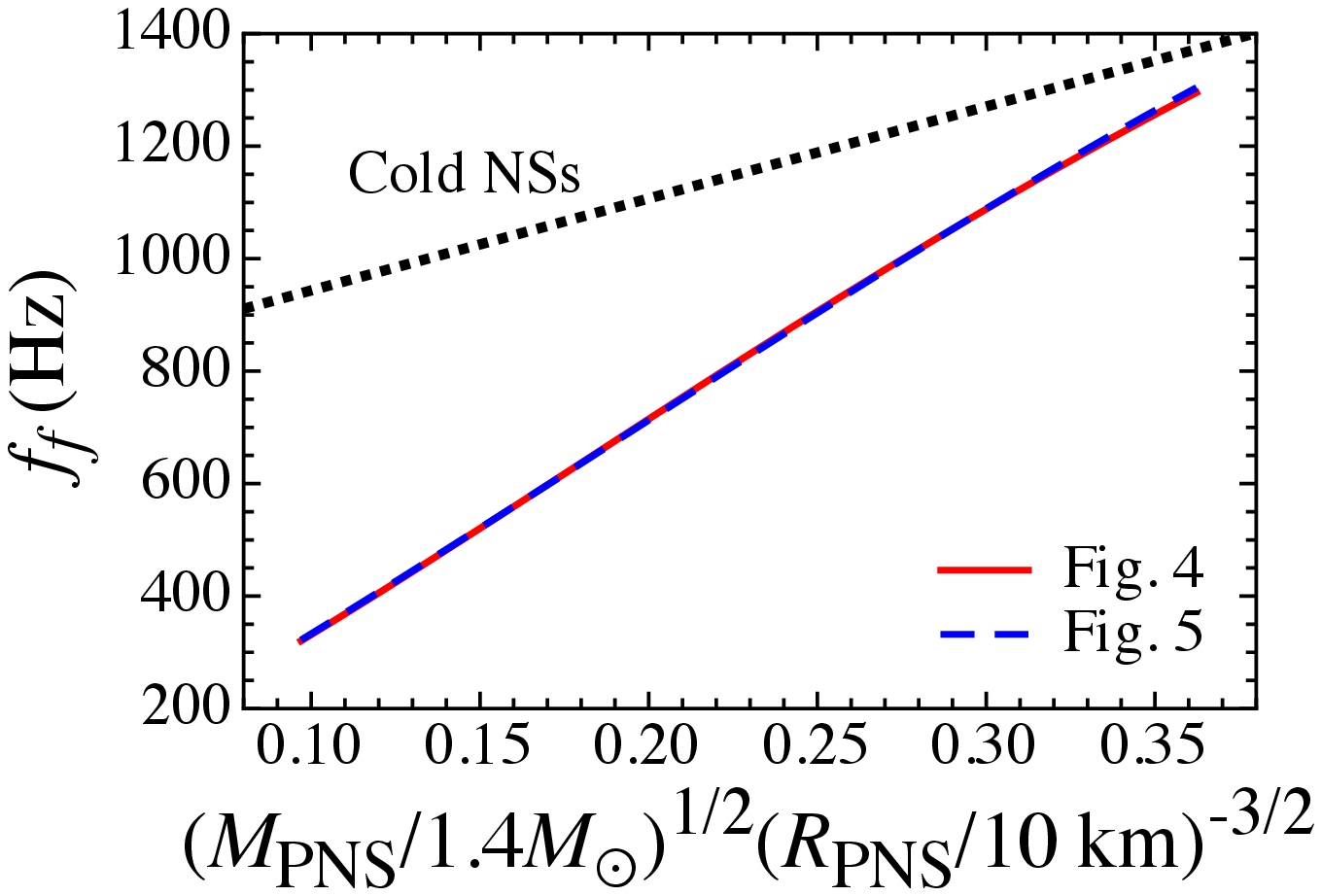} &
\includegraphics[scale=0.42]{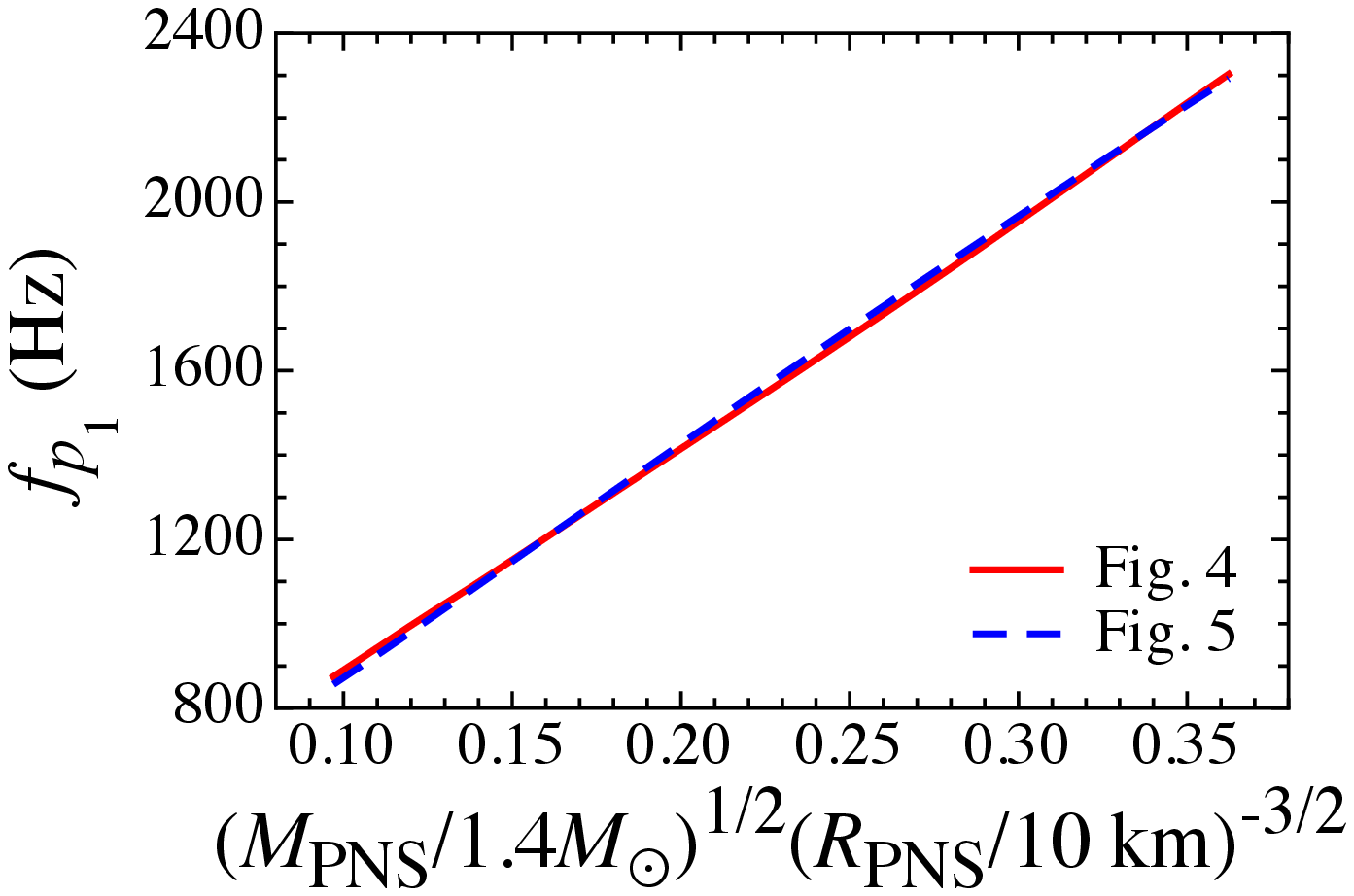} &
\includegraphics[scale=0.42]{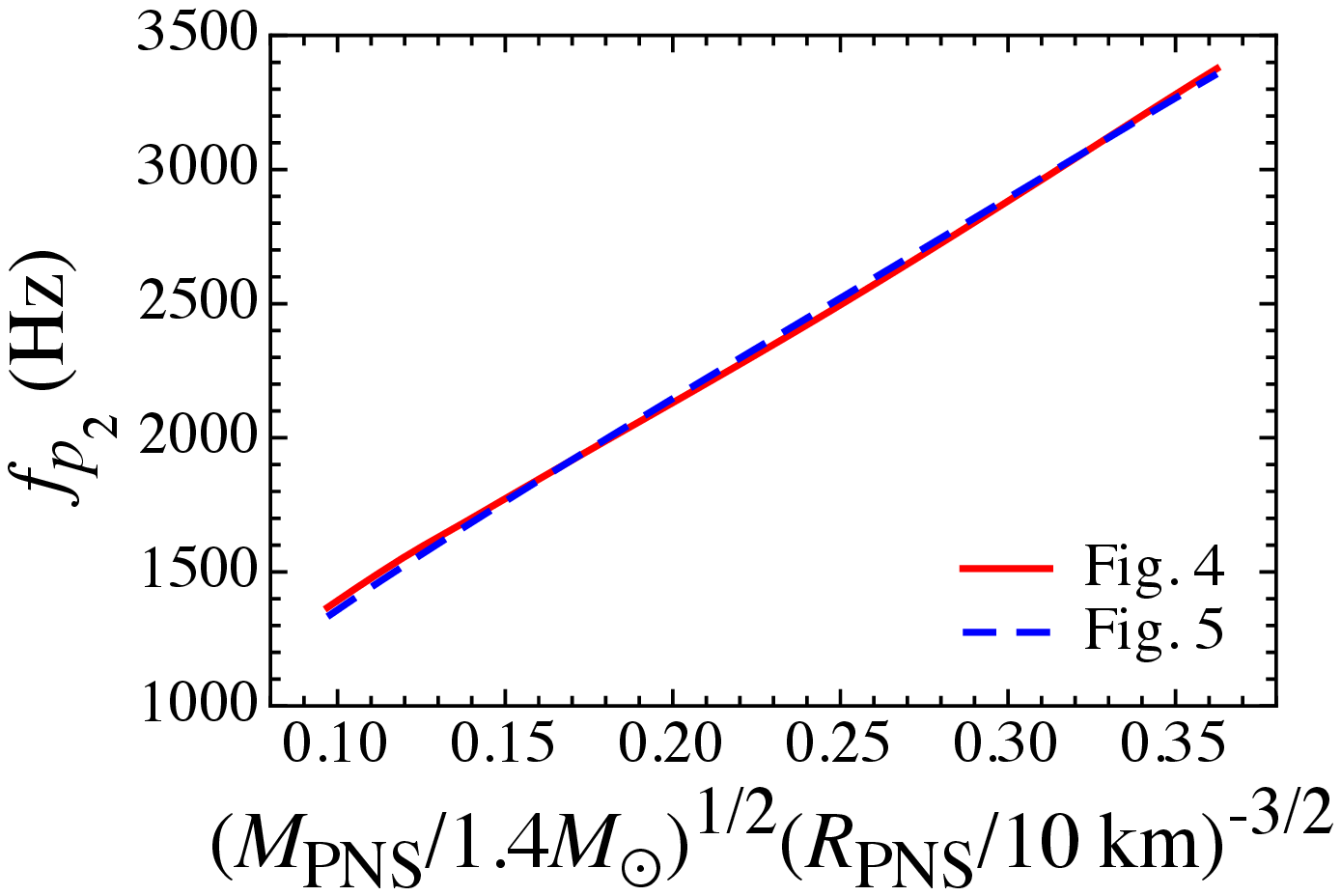}
\end{tabular}
\end{center}
\caption{
The frequencies of $f$-, $p_1$-, and $p_2$-modes are shown as a function of the normalized square root of the average density of PNSs for $M_{\rm pro}=15M_\odot$ and LS220, where the normalized square root of the average density is defined by $(M_{\rm PNS}/1.4M_\odot)^{1/2}(R_{\rm PNS}/10 {\rm km})^{-3/2}$. In the left panel, the $f$-mode frequencies expected for cold neutron stars with the same square root of the average density, which are given by Eq. (\ref{eq:AK}), are shown with the dotted line. 
}
\label{fig:f-ave-LS220M15}
\end{figure*}

In addition, since the $f$-, $p_1$-, and $p_2$-modes oscillations are acoustic waves, at least for cold neutron stars, it is known that the frequencies of such oscillations can be characterized by the square root of the average density of the star, which is defined as $(M/R^3)^{1/2}$ with mass ($M$) and radius ($R$) of cold neutron stars \cite{AK1998}. In fact, in Ref. \cite{AK1998}, they found that the $f$-mode frequencies can be expressed as 
\begin{equation}
  f_f^{\rm (NS)} {\rm \ (kHz)} \approx 0.78 + 1.635\left(\frac{M}{1.4M_\odot}\right)^{1/2}\left(\frac{R}{10\ {\rm km}}\right)^{-3/2},
  \label{eq:AK}
\end{equation}
which is almost independent from the EOS with which the neutron star model is constructed. So, we check the frequencies for the case of the PNSs. In Fig. \ref{fig:f-ave-LS220M15}, for the PNS model with $M_{\rm pro}=15M_\odot$ and LS220,  we show the frequencies of $f$-, $p_1$-, and $p_2$-modes as a function of the normalized square root of the average density, which is defined as $(M_{\rm PNS}/1.4M_\odot)^{1/2}(R_{\rm PNS}/10{\rm km})^{-3/2}$. Then, we find that the frequencies of $f$-, $p_1$-, and $p_2$-modes are proportional to the square root of the average density even for PNSs. We remark that, since PNS models become massive with small radius with time as shown in Fig. \ref{fig:MRLS220M15}, the PNS whose square root of the average density is high corresponds to the PNS model with time. Furthermore, in the left panel of Fig. \ref{fig:f-ave-LS220M15}, we also plot the $f$-mode frequencies expected for cold neutron stars, which are calculated with Eq. (\ref{eq:AK}), with the dotted line. From this panel, one can see that the $f$-mode frequencies for cold neutron stars obviously deviate from those for PNSs. In particular, such a deviation in $f$-mode frequencies becomes significantly for the stellar model with low square root of the average density. This deviation could come from the evidence that the stellar structure of PNS is different from that of cold neutron star, where the EOS for constructing the stellar model is different from each other.

\begin{figure*}
\begin{center}
\begin{tabular}{ccc}
\includegraphics[scale=0.42]{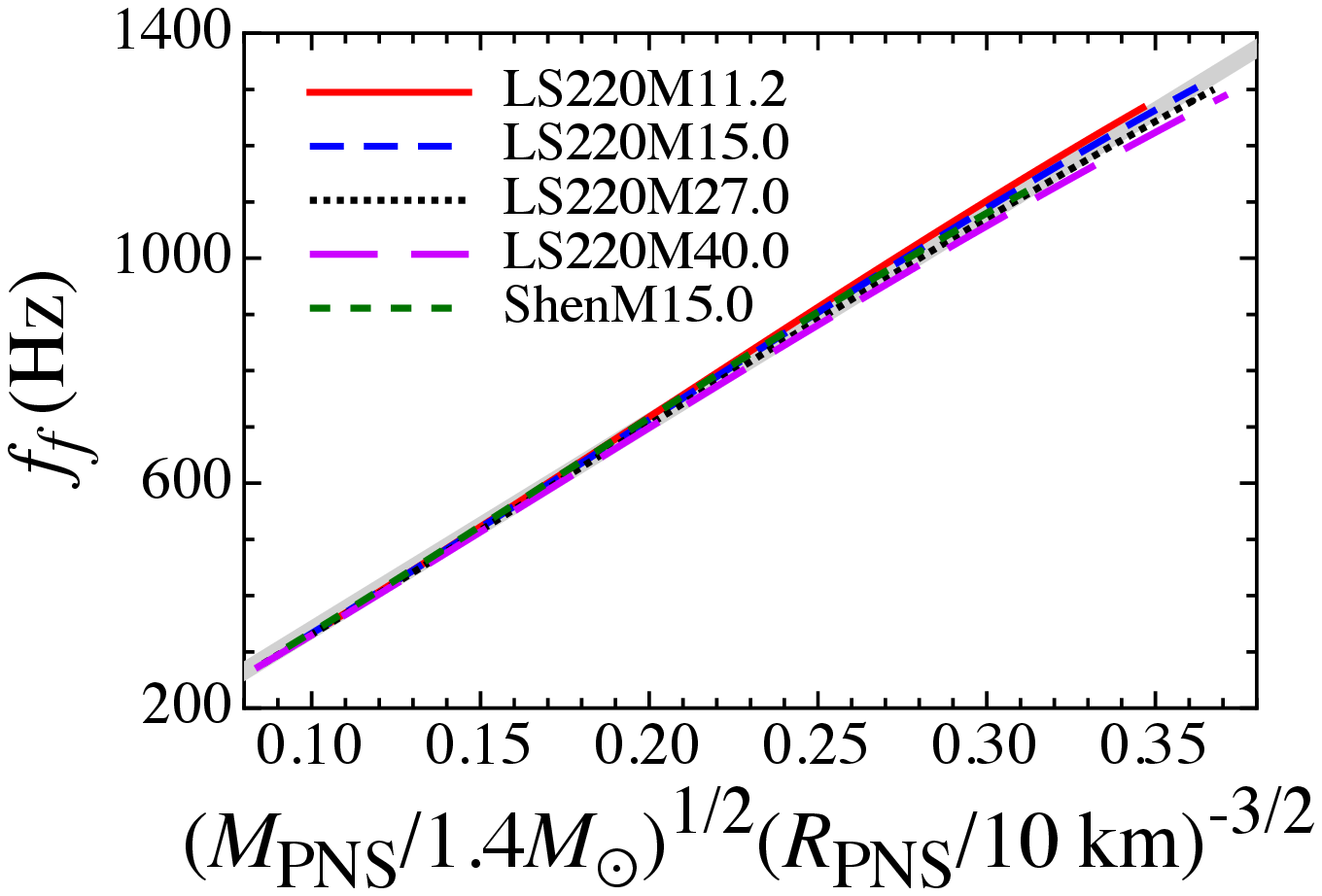} &
\includegraphics[scale=0.42]{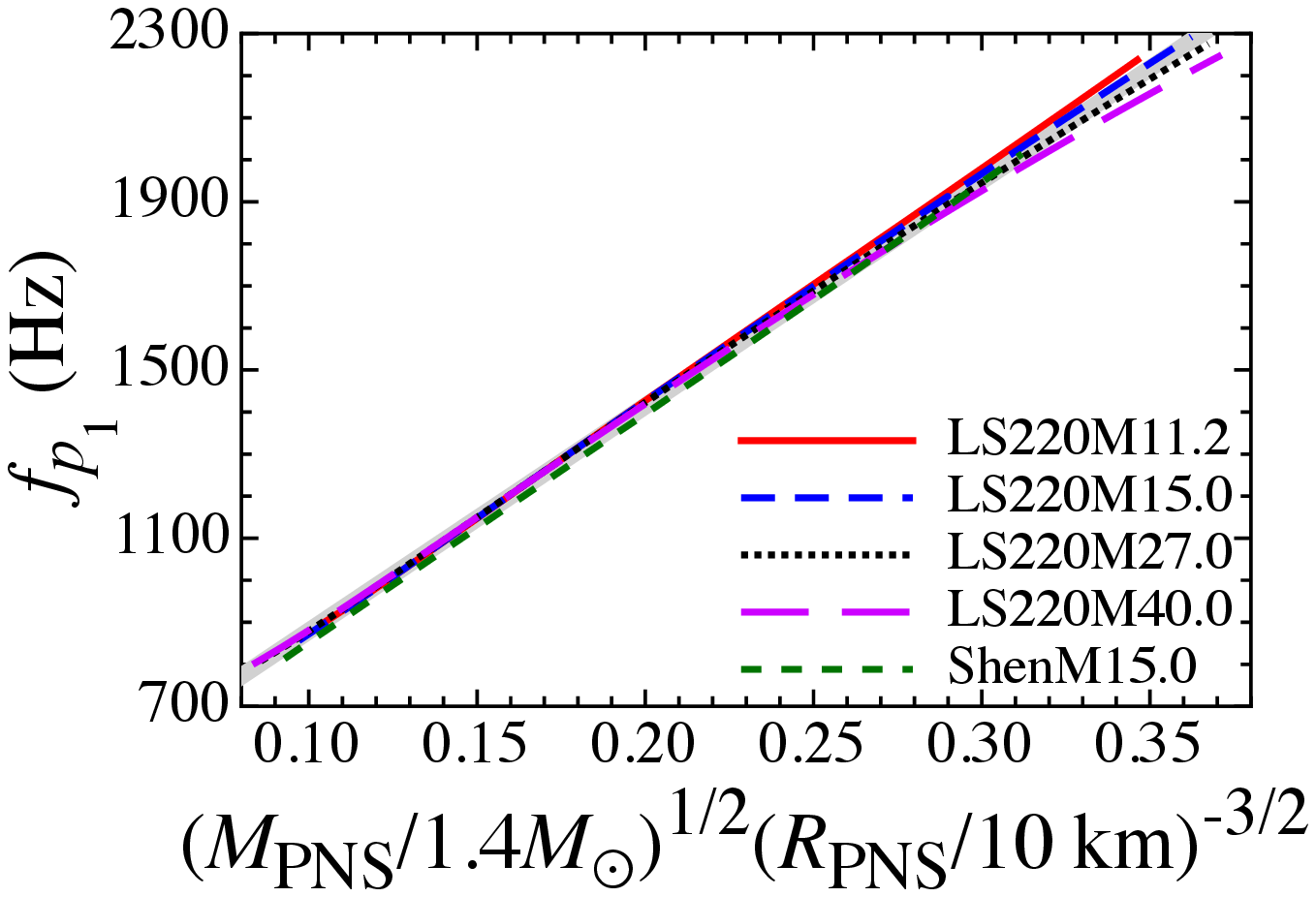} &
\includegraphics[scale=0.42]{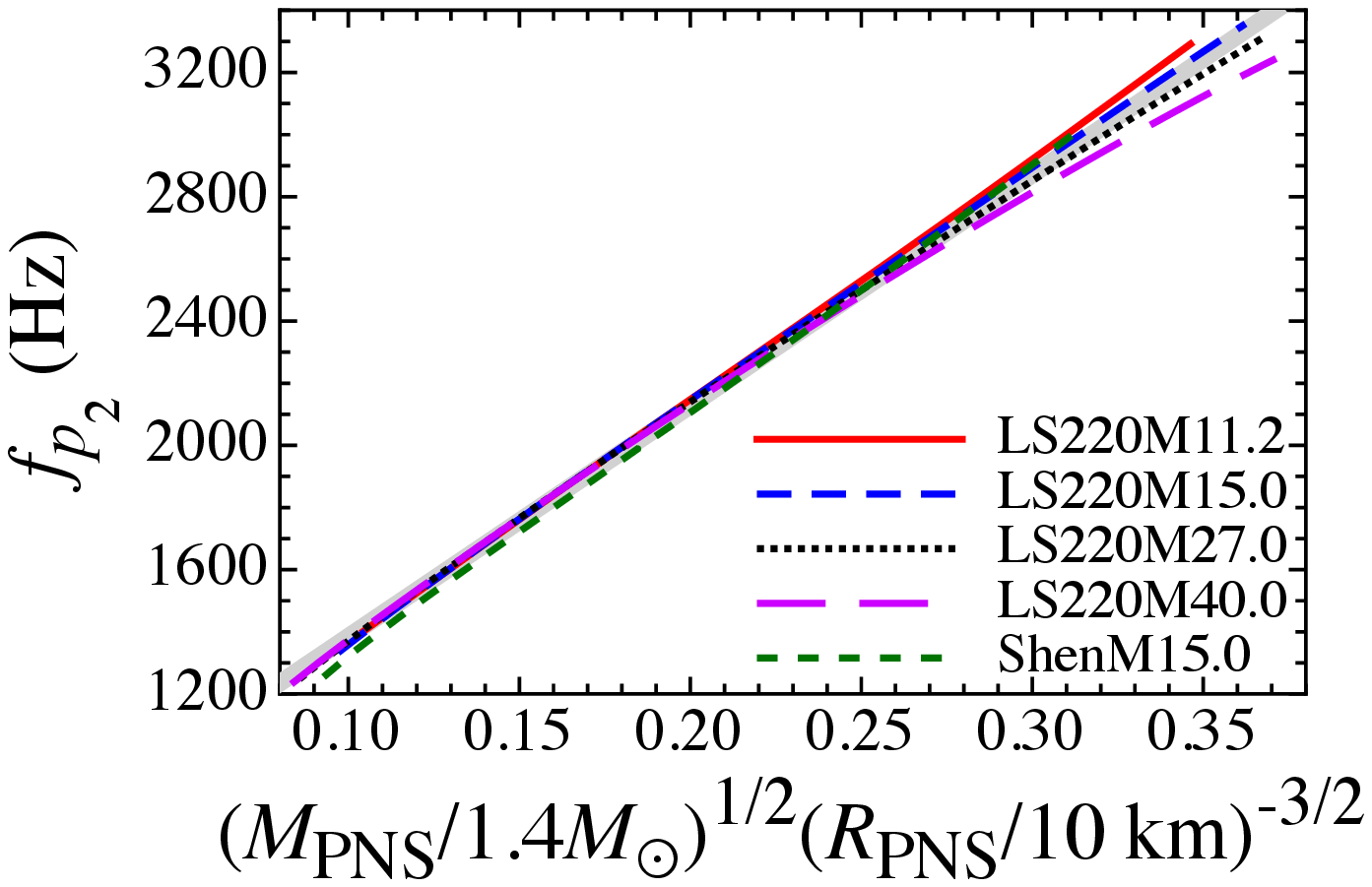}
\end{tabular}
\end{center}
\caption{
For the various progenitor models, the frequencies of $f$-, $p_1$-, and $p_2$-modes are shown as a function of the normalized square root of the average density of PNSs, where the normalized square root of the average density is defined by $(M_{\rm PNS}/1.4M_\odot)^{1/2}(R_{\rm PNS}/10 {\rm km})^{-3/2}$. The thick solid line in each panel corresponds to the universal relation shown as Eq. (\ref{eq:ff}).
}
\label{fig:f-ave-all}
\end{figure*}

\begin{figure*}
\begin{center}
\begin{tabular}{ccc}
\includegraphics[scale=0.42]{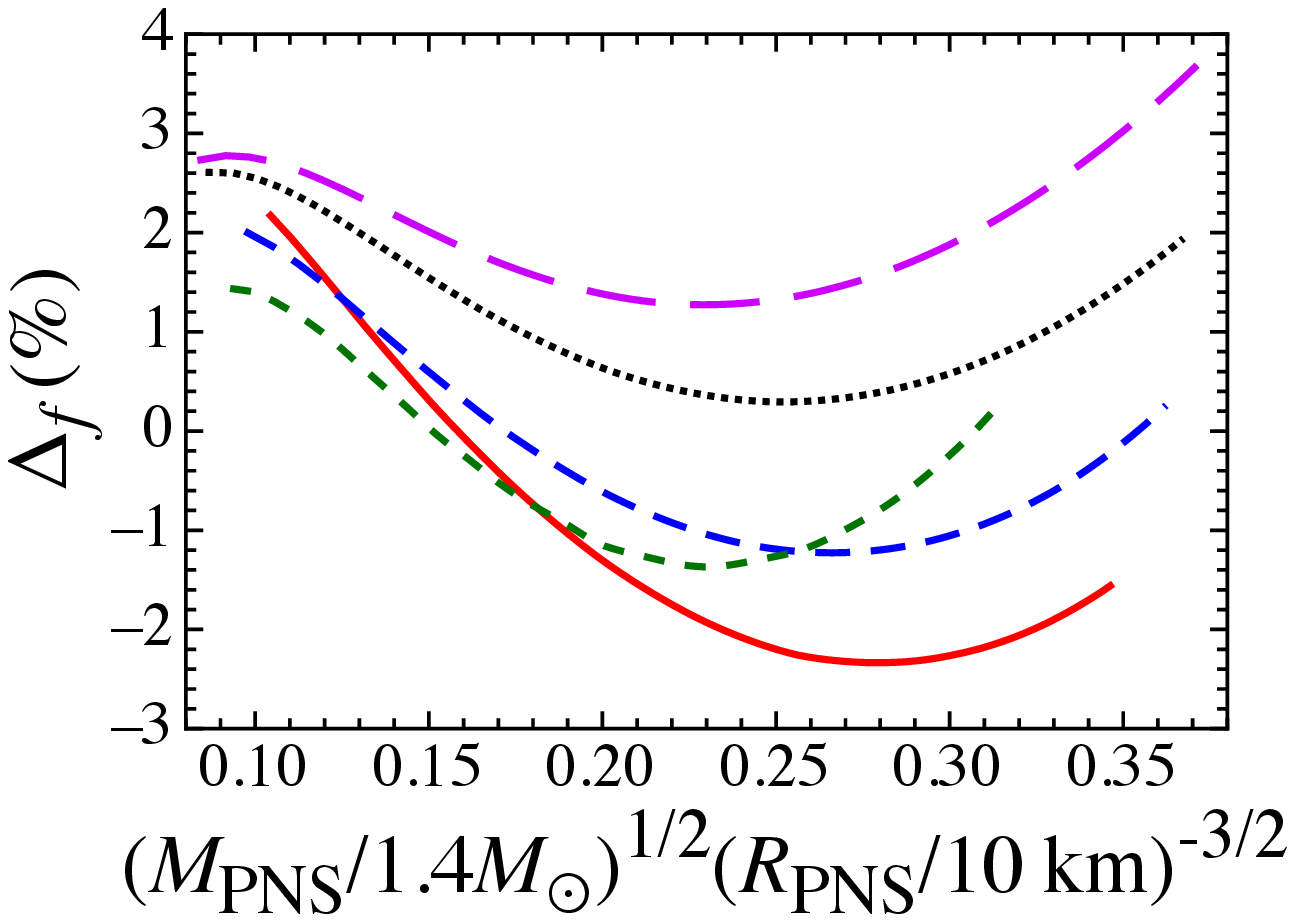} &
\includegraphics[scale=0.42]{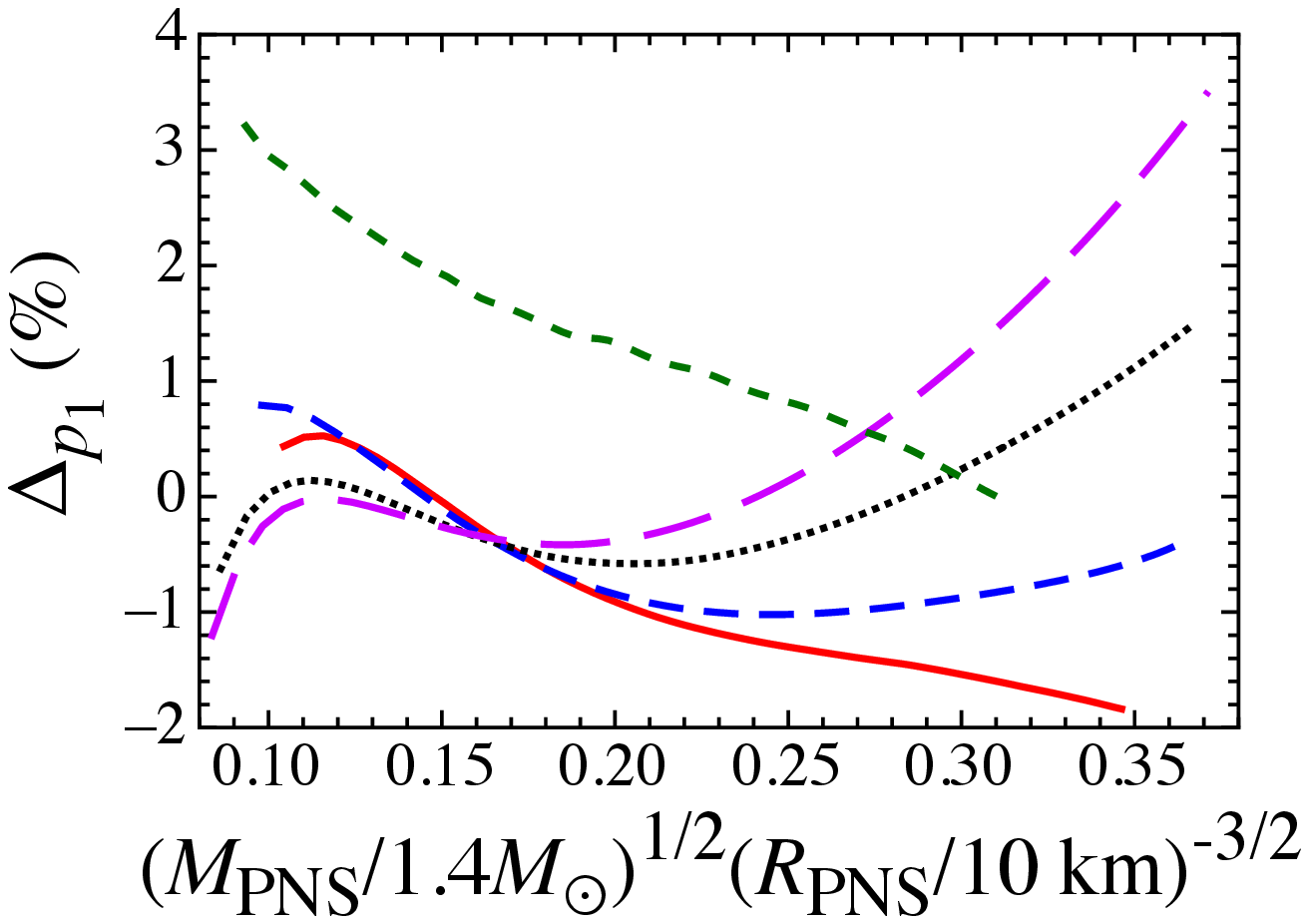} &
\includegraphics[scale=0.42]{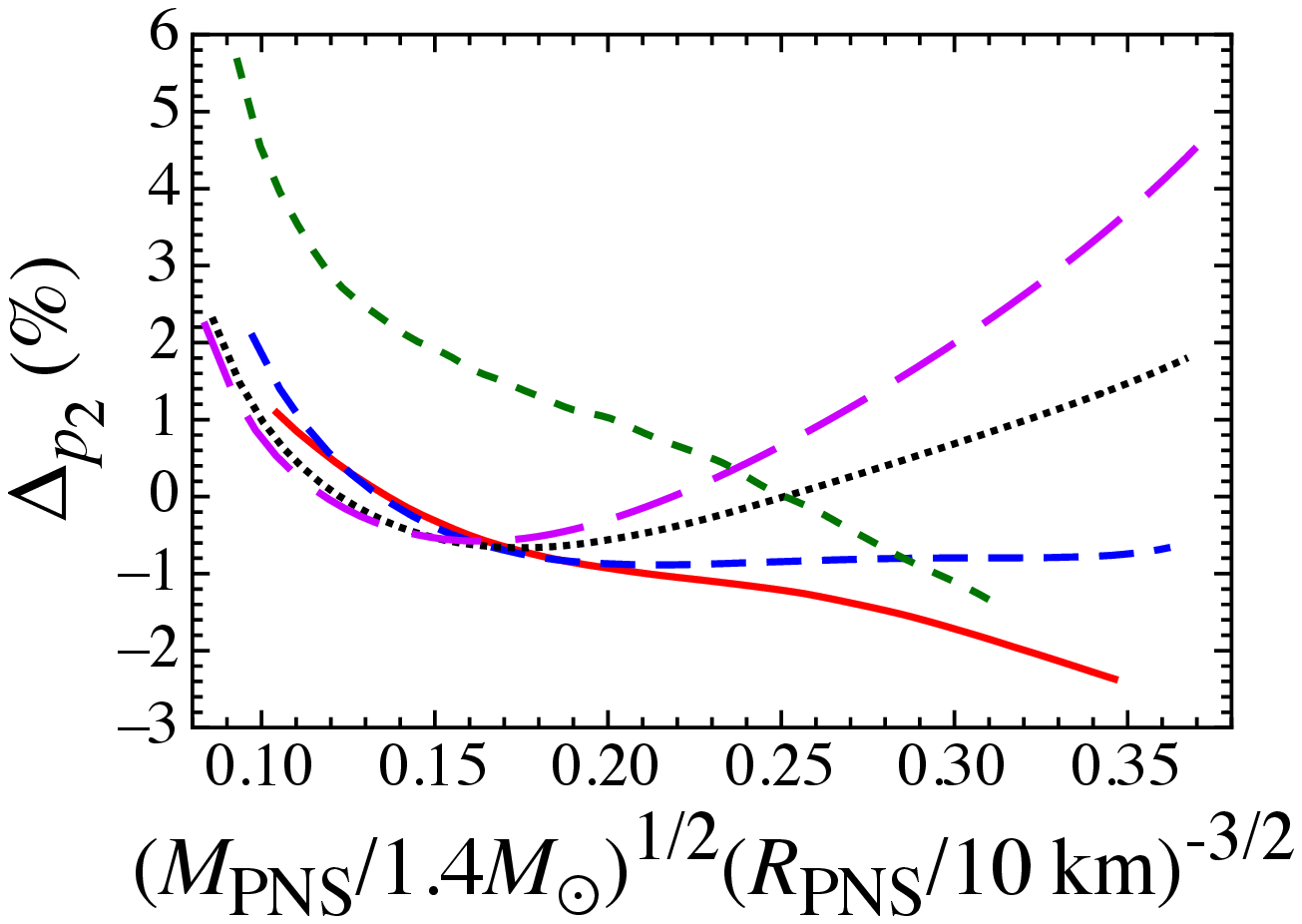}
\end{tabular}
\end{center}
\caption{
Relative deviation of the eigen-frequencies for each PNS model constructed with the different EOS from the universal relation given by Eq. (\ref{eq:ff}). The left, middle, and right panels correspond to the relative deviation for $f$-, $p_1$-, and $p_2$-modes. In figure, the meaning of lines is the same as in Fig. \ref{fig:f-ave-all}.
}
\label{fig:devi}
\end{figure*}

We also examine the frequencies of $f$-, $p_1$-, and $p_2$-modes for the various progenitor models of PNSs shown in Table \ref{tab:Rt}. We remark that we stop the calculations at 940 msec for the PNS constructed with $M_{\rm pro}=40.0M_\odot$ for LS220 EOS, because the PNS mass becomes more than the maximum mass  expected with LS220 around $950$ msec. Then, the obtained frequencies of $f$-, $p_1$, and $p_2$-modes for the various progenitor models are shown in Fig. \ref{fig:f-ave-all}, where the frequencies are calculated with the ($Y_e$, $s$) distributions inside the star as in Fig. \ref{fig:mYe2}. In this figure, LS220M11.2, LS220M15.0, LS220M27.0, and LS220M40.0 correspond to the results obtained with the progenitor models with $M_{\rm pro}=11.2M_\odot$, $15.0M_\odot$, $27.0M_\odot$, and $40.0M_\odot$ for LS220 EOS, respectively, while ShenM15.0 is the results obtained with the progenitor model with $M_{\rm pro}=15.0M_\odot$ for Shen EOS. From this figure, one can observe that the frequencies of PNSs are almost on the same line as a function of the square root of the average density of PNS, i.e., the frequencies are almost independent from the progenitor models. Thus, we can get an universal relation between the frequencies from the PNSs and the square root of the average density of PNSs, such as
\begin{equation}
   f_{i}^{\rm (PNS)} {\rm (Hz)} \approx c_i^0 + c_i^1\left(\frac{M_{\rm PNS}}{1.4M_\odot}\right)^{1/2}\left(\frac{R_{\rm PNS}}{10\ {\rm km}}\right)^{-3/2}, \label{eq:ff}
\end{equation}
where $i$ denotes $f$, $p_1$, and $p_2$ for $f$-, $p_1$, and $p_2$-modes, and $c_i^0$ and $c_i^1$ are some constants irrespective of the progenitor models of PNSs. The coefficients in this relation are shown in Table \ref{tab:ci01} and the universal relations obtained here are also plotted in Fig. \ref{fig:f-ave-all} with thick solid line. Additionally, the relative deviation of the eigen-frequencies for the PNS models constructed with various progenitor models from the expectation with Eq. (\ref{eq:ff}) are shown in Fig. \ref{fig:devi}. The relative deviation $\Delta_i$ is calculated by
\begin{equation}
  \Delta_i = \frac{\left[f_i^{\rm (PNS)} - f_i\right]}{f_i^{\rm (PNS)}}\times 100,
\end{equation} 
where $f_i^{\rm (PNS)}$ and $f_i$ are the frequencies given by Eq. (\ref{eq:ff}) and those obtained by numerical calculation for each PNS model with various progenitor model. From this figure, one can see that the relative deviation from the universal relation becomes larger for higher overtones.



\begin{table}[htbp]
\begin{center}
\leavevmode
\caption{Coefficients in the universal relation shown as Eq. (\ref{eq:ff}) for the various progenitor models of PNSs.
}
\begin{tabular}{ccccc}
\hline\hline
  & modes & $c_i^0$ (Hz) & $c_i^1$ (Hz) &  \\
\hline
  & $f$      & $-29.48$ & $3690$ & \\
  & $p_1$ & $343.9$ & $5352$ & \\
  & $p_2$ & $640.8$ & $7435$ & \\
\hline\hline
\end{tabular}
\label{tab:ci01}
\end{center}
\end{table}

\begin{figure}
\begin{center}
\includegraphics[scale=0.5]{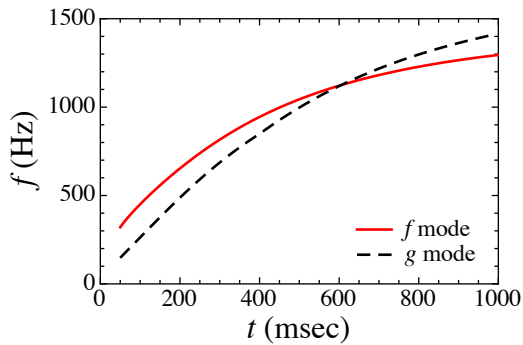}
\end{center}
\caption{
Time evolution of frequencies of the $f$- and $g$-modes radiating from PNSs for $M_{\rm pro}=15M_\odot$ and LS220. 
}
\label{fig:Mueller}
\end{figure}

With respect to the characteristic gravitational waves radiating after bounce of core-collapse supernovae, the evidence of signal due to the convection and the standing accretion-shock instability has also been reported \cite{MJM2013,Pablo2013,Yakunin2015,KKT2016,AMMJ2016}, which is associated with the $g$-mode oscillations around (and above) the surface of PNSs. In fact, the frequencies can be well-expressed by using the radius and mass of PNSs as
\begin{equation}
   f_{g} \approx \frac{1}{2\pi} \frac{GM_{\rm PNS}}{R_{\rm PNS}^2} \left(\frac{1.1m_n}{\langle E_{\bar{\nu}_e}\rangle }\right)^{1/2} \left(1-\frac{GM_{\rm PNS}}{c^2R_{\rm PNS}}\right)^2, \label{eq:fg}
\end{equation}
where $m_n$ and $\langle E_{\bar{\nu}_e}\rangle$ denote the neutron mass and the mean energy of electron anti-neutrinos \cite{MJM2013}. That is, the frequencies essentially depend on $M_{\rm PNS}/R_{\rm PNS}^2$, which is completely different from the $f$-mode frequencies depending on $(M_{\rm PNS}/R_{\rm PNS}^3)^{1/2}$ as shown above. Thus, carefully observing the frequencies of gravitational waves radiating from the PNSs in supernovae, one might be possible to determine the mass and radius of PNSs via Eqs. (\ref{eq:ff}) and (\ref{eq:fg}) \cite{distance}. For example, one might observe the time evolution of gravitational wave spectra from the PNS for $M_{\rm pro}=15M_\odot$ and LS220, as shown in Fig. \ref{fig:Mueller}. We remark that, to calculate the $g$-mode frequencies with Eq. (\ref{eq:fg}), we adopt the $\langle E_{\bar{\nu}_e}\rangle$ distribution given by 
\begin{equation}
  \langle E_{\bar{\nu}_e}\rangle =   \left\{
    \begin{array}{ll}
      3t/400 + 13 &(0\le t \le 400\ {\rm msec}) \\
      16 &(400\ {\rm msec}\le t) \ 
    \end{array}
  \right.,\label{eq:Enu}
\end{equation}
which is an imitation of Fig. 1 in Ref. \cite{MJ2014}, and the radius and mass of PNS are determined so that the surface density sets to be $10^{11}$ g/cm$^3$ as in Ref. \cite{MJM2013}.

\section{Conclusion}
\label{sec:IV}

In this paper, we consider the time evolution of the gravitational wave spectra radiating from the PNSs after bounce in core-collapse supernova. For this purpose, we construct PNS models in such a way that the mass and radius obtained from the 1D simulation are mimicked under the assumption that the PNS models are hydrostatic on each time step. On such PNS models, we calculate the eigen frequencies by solving the eigenvalue problem within the relativistic linear analysis. Then, we find that the frequencies from the PNSs are almost independent from the distributions of electron fraction and of the entropy per baryon inside the stars. In addition, the frequencies can be characterized by the square root of the average density of PNSs, which are almost independent from the progenitor models. Eventually, we can derive the universal relation between the frequencies from PNSs and square root of the average density of PNSs. Since this relation is completely different from the empirical relation for the frequencies due to the convection and the standing accretion-shock instability around the surface of PNS, one might be possible to determine the mass and radius of PNS via careful observations of gravitational waves after core-collapse supernova. In this paper, as a first step, we construct PNS models with the 1D simulation of core-collapse supernovae within Newtonian theory. So, we will perform the similar analysis for the PNS models with higher dimension simulation and/or within relativistic framework in future.

\acknowledgments
This work was supported in part by Grant-in-Aid for Young Scientists (B) through Grant Nos. 26800133 (H.S.) and 26870823 (T.T.) provided by JSPS and by Grants-in-Aid for Scientific Research on Innovative Areas through Grant Nos. 15H00843 (H.S.),  15H01039 (T.T.), and 15H00789 (T.T.) provided by MEXT. T.T. was supported by MEXT and JICFuS as a priority issue (Elucidation of the fundamental laws and evolution of the universe) to be tackled by using Post `K' Computer. The computations in this research were performed by PC cluster at Center for Computational Astrophysics, National Astronomical Observatory of Japan.





\begin{thebibliography}{999}

\bibitem{Haen2007}
P. {Haensel}, A. {Potekhin}, and D. G. {Yakovlev},
\newblock {\em {Neutron stars I}} (Springer, 2007).

\bibitem{D2010}
P. B. {Demorest}, T. {Pennucci}, S. M. {Ransom}, M. S. E. {Roberts}, and J. W. T. {Hessels},
\newblock Nature {\bf 467}, 1081 (2010), arXiv:1010.5788.

\bibitem{A2013}
J. {Antoniadis} {\em et al.},
\newblock Science {\bf 340}, 448 (2013), arXiv:1304.6875.

\bibitem{AK1996}
N. {Andersson} and K. D. {Kokkotas},
\newblock Phys. Rev. Lett. {\bf 77}, 4134 (1996), gr-qc/9610035.

\bibitem{AK1998}
N. {Andersson} and K. D. {Kokkotas},
\newblock Mon. Not. R. Astron. Soc. {\bf 299}, 1059 (1998), gr-qc/9711088.

\bibitem{STM2002}
H. {Sotani}, K. {Tominaga}, and K.-I. {Maeda},
\newblock \prd {\bf 65}, 024010 (2001), gr-qc/0108060.

\bibitem{SH2003}
H. {Sotani} and T. {Harada},
\newblock \prd {\bf 68}, 024019 (2003), gr-qc/0307035.

\bibitem{SYMT2011}
H. {Sotani}, N. {Yasutake}, T. {Maruyama}, and T. {Tatsumi},
\newblock \prd {\bf 83}, 024014 (2011), arXiv:1012.4042.

\bibitem{PA2012}
A. {Passamonti} and N. {Andersson},
\newblock Mon. Not. R. Astron. Soc. {\bf 419}, 638 (2012), arXiv:1105.4787.

\bibitem{DGKK2013}
D. D. {Doneva}, E. {Gaertig}, K. D. {Kokkotas}, and C. {Kr{\"u}ger},
\newblock \prd {\bf 88}, 044052 (2013), arXiv:1305.7197.

\bibitem{WS2006}
A. L. Watts and T. E. Strohmayer,
\newblock Advances Space Res. {\bf 40}, 1446 (2006).

\bibitem{SW2009}
A. W. {Steiner} and A. L. {Watts},
\newblock Phys. Rev. Lett. {\bf 103}, 181101 (2009), arXiv:0902.1683.

\bibitem{GNJL2011}
M. {Gearheart}, W. G. {Newton}, J. {Hooker}, and B.-A. {Li},
\newblock Mon. Not. R. Astron. Soc. {\bf 418}, 2343 (2011), arXiv:1106.4875.

\bibitem{SNIO2012}
H. {Sotani}, K. {Nakazato}, K. {Iida}, and K. {Oyamatsu},
\newblock Phys. Rev. Lett. {\bf 108}, 201101 (2012), arXiv:1202.6242.

\bibitem{SNIO2013a}
H. {Sotani}, K. {Nakazato}, K. {Iida}, and K. {Oyamatsu},
\newblock Mon. Not. R. Astron. Soc. {\bf 428}, L21 (2013), arXiv:1210.0955.

\bibitem{SNIO2013b}
H. {Sotani}, K. {Nakazato}, K. {Iida}, and K. {Oyamatsu},
\newblock Mon. Not. R. Astron. Soc. {\bf 434}, 2060 (2013), arXiv:1303.4500.

\bibitem{SIO2016}
H. {Sotani}, K. {Iida}, and K. {Oyamatsu},
\newblock New Astronomy {\bf 43}, 80 (2016), arXiv:1508.01728.

\bibitem{Ferrari:2002ut}
V. Ferrari, G. Miniutti, and J. A. Pons,
\newblock Mon. Not. Roy. Astron. Soc. {\bf 342}, 629 (2003),
  arXiv:astro-ph/0210581.

\bibitem{Pons:1998mm}
J. A. Pons, S. Reddy, M. Prakash, J. M. Lattimer, and J. A. Miralles,
\newblock Astrophys. J. {\bf 513}, 780 (1999), arXiv:astro-ph/9807040.

\bibitem{Pons:2001ar}
J. A. Pons, A. W. Steiner, M. Prakash, and J. M. Lattimer,
\newblock Phys. Rev. Lett. {\bf 86}, 5223 (2001), arXiv:astro-ph/0102015.

\bibitem{LIGO}
G. M. Harry and the LIGO Scientific Collaboration,
\newblock Classical and Quantum Gravity {\bf 27}, 084006 (2010).

\bibitem{VIRGO}
S. Hild {\em et al.},
\newblock Classical and Quantum Gravity {\bf 26}, 025005 (2009).

\bibitem{KAGRA}
K. Somiya,
\newblock Classical and Quantum Gravity {\bf 29}, 124007 (2012).

\bibitem{OOGAGS2013}
C. D. {Ott} {\em et al.},
\newblock Nuclear Physics B Proceedings Supplements {\bf 235}, 381 (2013),
  arXiv:1212.4250.

\bibitem{GSSZGO2016}
S. E. {Gossan}, P. Sutton, A. Stuver, M. Zanolin, K. Gill, and C. D. Ott,
\newblock \prd {\bf 93}, 042002 (2016), arXiv:1511.02836.

\bibitem{NHTHTK2016}
K. {Nakamura} {\em et al.},
\newblock ArXiv e-prints  (2016), arXiv:1602.03028.

\bibitem{HKKT2015}
K. {Hayama}, T. {Kuroda}, K. {Kotake}, and T. {Takiwaki},
\newblock \prd {\bf 92}, 122001 (2015), arXiv:1501.00966.

\bibitem{TargetedSearch}
B. P. {Abbott} {\em et al.},
\newblock ArXiv e-prints  (2016), arXiv:1605.01785.

\bibitem{thierry15}
T. {Foglizzo} {\em et al.},
\newblock \pasa {\bf 32}, e009 (2015), arXiv:1501.01334.

\bibitem{tony15}
A. {Mezzacappa} {\em et al.},
\newblock ArXiv e-prints  (2015), arXiv:1501.01688.

\bibitem{JMS2016}
H.-T. {Janka}, T. {Melson}, and A. {Summa},
\newblock ArXiv e-prints  (2016), arXiv:1602.05576.

\bibitem{burrows13}
A. {Burrows},
\newblock Reviews of Modern Physics {\bf 85}, 245 (2013), arXiv:1210.4921.

\bibitem{Kotake12_ptep}
K. {Kotake} {\em et al.},
\newblock Progress of Theoretical and Experimental Physics {\bf 2012}, 010000
  (2012), arXiv:1205.6284.

\bibitem{MOB2009}
J. W. {Murphy}, C. D. {Ott}, and A. {Burrows},
\newblock Astrophys. J. {\bf 707}, 1173 (2009), arXiv:0907.4762.

\bibitem{MJM2013}
B. {M{\"u}ller}, H.-T. {Janka}, and A. {Marek},
\newblock Astrophys. J. {\bf 766}, 43 (2013), arXiv:1210.6984.

\bibitem{Pablo2013}
P. {Cerd{\'a}-Dur{\'a}n}, N. {DeBrye}, M. A. {Aloy}, J. A. {Font}, and
  M. {Obergaulinger},
\newblock Astrophys. J. Lett. {\bf 779}, L18 (2013), arXiv:1310.8290.

\bibitem{YMMYBHLBEVL2016}
   K. N. {Yakunin}, A. Mezzacappa, P. Marronetti, S. Yoshida, S. W. Bruenn, W. R. Hix,
   E. J. Lentz, O. E. Bronson Messer, J. A. Harris, E. Endeve, J. M. Blondin, and E. J. Lingerfelt,
\newblock \prd {\bf 92}, 084040 (2015), arXiv:1505.05824.

\bibitem{DOJMM2007}
H. {Dimmelmeier}, C. D. {Ott}, H.-T. {Janka}, A. {Marek}, and E. {M{\"u}ller},
\newblock Phys. Rev. Lett. {\bf 98}, 251101 (2007), astro-ph/0702305.

\bibitem{HDKMRSY2008}
K. {Hayama} {\em et al.},
\newblock Classical and Quantum Gravity {\bf 25}, 184022 (2008),
  arXiv:0807.4514.

\bibitem{SBFO2008}
T. Z. {Summerscales}, A. {Burrows}, L. S. {Finn}, and C. D. {Ott},
\newblock Astrophys. J. {\bf 678}, 1142 (2008), arXiv:0704.2157.

\bibitem{RBCDHM2009}
C. {R{\"o}ver}, M. A. Bizouard, N. Christensen, H. Dimmelmeier, I. S. Heng, and R. Meyer,
\newblock \prd {\bf 80}, 102004 (2009), arXiv:0909.1093.

\bibitem{LOHKS2012}
J. {Logue}, C. D. {Ott}, I. S. {Heng}, P. {Kalmus}, and J. H. C. {Scargill},
\newblock \prd {\bf 86}, 044023 (2012), arXiv:1202.3256.

\bibitem{AGDO2014}
E. {Abdikamalov}, S. {Gossan}, A. M. {DeMaio}, and C. D. {Ott},
\newblock \prd {\bf 90}, 044001 (2014), arXiv:1311.3678.

\bibitem{EMC2014}
M. C. {Edwards}, R. {Meyer}, and N. {Christensen},
\newblock Inverse Problems {\bf 30}, 114008 (2014), arXiv:1407.7549.

\bibitem{EFO2014}
W. J. {Engels}, R. {Frey}, and C. D. {Ott},
\newblock \prd {\bf 90}, 124026 (2014), arXiv:1406.1164.

\bibitem{YAKSKKV2015}
T. {Yokozawa} {\em et al.},
\newblock Astrophys. J. {\bf 811}, 86 (2015), arXiv:1410.2050.

\bibitem{Langer2012}
N. {Langer},
\newblock \araa {\bf 50}, 107 (2012), arXiv:1206.5443.

\bibitem{SKJM2006}
L. {Scheck}, K. {Kifonidis}, H.-T. {Janka}, and E. {M{\"u}ller},
\newblock \aap {\bf 457}, 963 (2006), astro-ph/0601302.

\bibitem{Roberts2012}
L. F. {Roberts},
\newblock Astrophys. J. {\bf 755}, 126 (2012), arXiv:1205.3228.

\bibitem{LS}
J. M. {Lattimer} and F. {Douglas Swesty},
\newblock Nuclear Physics A {\bf 535}, 331 (1991).

\bibitem{Shen}
H. {Shen}, H. {Toki}, K. {Oyamatsu}, and K. {Sumiyoshi},
\newblock Nuclear Physics A {\bf 637}, 435 (1998), arXiv:nucl-th/9805035.

\bibitem{WW1995}
S. E. {Woosley} and T. A. {Weaver},
\newblock \apjs {\bf 101}, 181 (1995).

\bibitem{WHW2002}
S. E. {Woosley}, A. {Heger}, and T. A. {Weaver},
\newblock Reviews of Modern Physics {\bf 74}, 1015 (2002).

\bibitem{TKS2016}
T. {Takiwaki}, K. {Kotake}, and Y. {Suwa},
\newblock ArXiv e-prints  (2016), arXiv:1602.06759.

\bibitem{NTKT2015}
K. {Nakamura}, T. {Takiwaki}, T. {Kuroda}, and K. {Kotake},
\newblock \pasj {\bf 67}, 107 (2015), arXiv:1406.2415.

\bibitem{HLLE}
B. {Einfeldt},
\newblock SIAM Journal on Numerical Analysis {\bf 25}, 294 (1988).

\bibitem{idsa}
M. {Liebend{\"o}rfer}, S. C. {Whitehouse}, and T. {Fischer},
\newblock Astrophys. J. {\bf 698}, 1174 (2009), arXiv:0711.2929.

\bibitem{Bruenn85}
S. W. {Bruenn},
\newblock \apjs {\bf 58}, 771 (1985).

\bibitem{MB1993}
A. {Mezzacappa} and S. W. {Bruenn},
\newblock Astrophys. J. {\bf 410}, 740 (1993).

\bibitem{RJ2002}
M. {Rampp} and H.-T. {Janka},
\newblock \aap {\bf 396}, 361 (2002), astro-ph/0203101.

\bibitem{Horowitz1997}
C. J. {Horowitz},
\newblock \prd {\bf 55}, 4577 (1997), astro-ph/9603138.

\bibitem{HR1998}
S. {Hannestad} and G. {Raffelt},
\newblock Astrophys. J. {\bf 507}, 339 (1998), astro-ph/9711132.

\bibitem{M2010}
B. {M{\"u}ller}, H.-T. {Janka}, and H. {Dimmelmeier},
\newblock \apjs {\bf 189}, 104 (2010), arXiv:1001.4841.

\bibitem{entropy}
 The evolution of entropy profile is determined by the competition between the entropy increase via the release of gravitational binding energy after bounce and the decrease via the neutrino cooling due to the diffusion process. The neutrino opacity in the vicinity of stellar center is so high that the neutrino can not escape easily, while that for outer region of protoneutron star becomes low where the neutrino can escape efficiently. Such a neutrino opacity depending on the density leads to the evolution of entropy profile as shown in Fig. \ref{fig:Yes12}. In any case, the central profile of entropy after $\sim 400$ msec. deviates from the previous numerical results in general relativity (e.g., \cite{Roberts2012}), because our simulation has been done in Newtonian.

\bibitem{thermal}
 We distinguish the thermal energy from the entire internal energy in a sense that the thermal energy is just defined by the temperature, while the entire internal energy includes the chemical potential. 

\bibitem{Yakunin2015}
  K. N. Yakunin, A. Mezzacappa, P. Marronetti, S. I. Yoshida, S. W. Bruenn, W. Raphael Hix, 
  E. J. Lentz, O. E. Bronson Messer, J. Austin Harris, E. Endeve, J. M. Blondin, and E. J. Lingerfelt, 
  ArXiv e-prints  (2015), arXiv:1505.05824.

\bibitem{KKT2016}
  T. Kuroda, K. Kotake, and T. Takiwaki, ArXiv e-prints (2016), arXiv:1605.09215.

\bibitem{AMMJ2016}
  H. Andresen, B. M\"{u}ller, E. M\"{u}ller, and H.-T. Janka, ArXiv e-prints (2016), arXiv:1607.05199.

\bibitem{distance}
  According to the simulation by Andresen et al. \cite{AMMJ2016}, the gravitational waves due to the convection and the standing accretion-shock instability could be detected if the source is closer than $5-10$ kpc, depending on the progenitor mass. Thus, this value might be a critical distance for determining the PNS mass and radius via the direct observations of $f$ and $g$-modes oscillations.

\bibitem{MJ2014}
B. {M{\"u}ller} and H.-T. {Janka},
\newblock Astrophys. J. {\bf 788}, 82 (2014), arXiv:1402.3415.

\end{thebibliography}
\end{document}